\documentclass{jpp}
\usepackage{aas_macros}
\usepackage{graphicx}
\usepackage[justification=centering]{subcaption}
\usepackage[justification=centering]{caption}

\usepackage[utf8]{inputenc}
\usepackage[T1]{fontenc}
\usepackage{amsmath}

\usepackage{color}

\shorttitle{Solar wind variations over a cycle with asymmetry}
\shortauthor{B. Perri, A. S. Brun, V. Réville and  A. Strugarek }

\title{Simulations of solar wind variations during an 11-year cycle and the influence of north-south asymmetry}

\author{
  B. Perri\aff{1},
  A. S. Brun\aff{1},
  V. Réville\aff{2,1},
 \and A. Strugarek\aff{1}
}

\affiliation{
\aff{1}
AIM, CEA, CNRS, Université Paris-Saclay, Université Paris-Diderot, Sorbonne Paris Cité, F-91191 Gif-sur-Yvette, France
\aff{2}EPSS, University of California, Los Angeles, CA, United States
}

\begin{document}

\maketitle

\begin{abstract}
We want to study the connections between the magnetic field generated inside the Sun and the solar wind impacting Earth, especially the influence of north-south asymmetry on the magnetic and velocity fields. We study a solar-like 11-year cycle in a quasi-static way : an asymmetric dynamo field is generated through a 2.5D flux-transport model with Babcock-Leighton mechanism, and then is used as bottom boundary condition for compressible 2.5D simulations of the solar wind. We recover solar values for the mass loss rate, the spin-down timescale and the Alfvén radius, and are able to reproduce the observed delay in latitudinal variations of the wind and the general wind structure observed for the Sun. We show that the phase-lag between the energy of the dipole component and the total surface magnetic energy has a strong influence on the amplitude of the variations of global quantities. We show in particular that the magnetic torque variations can be linked to topological variations during a magnetic cycle, while variations in the mass loss rate appeared to be driven by variations of the magnetic energy. 
\end{abstract}

\section{Introduction}\label{sec:intro}

From a space weather perspective, one of the main challenges is to model accurately the solar wind, for it has a profound effect on the Earth's space environment. Various in-situ observation missions, like \textit{Ulysses} for example have shown that the magnetic activity has a strong influence on the wind structure and velocity, depending on the phase of the 11-year cycle (see \cite{McComas2008} and \cite{Smith2011} for summary of the mission highlights, and \cite{Issautier2008} for discussion about north-south asymmetry). At minimum of activity, the magnetic field is mostly dipolar and the wind is slower at the equator (around 400 km/s) and faster at the poles (around 800 km/s); at maximum of activity, the magnetic field is multipolar and the wind distribution is bimodal at all latitudes. This is why there is more and more effort from an instrumental and theoretical point of view, to link in-situ space measurements and remote sensing solar surface observations, one of the goals of ESA's Solar Orbiter mission.

Fortunately, we have a lot of observational data when it comes to the Sun magnetic activity : sunspots and magnetic field observations from several observatories and several time periods (Royal Greenwich Observatory from 1874 to 1954 in \cite{Newton1955} and from 1874 to 1976 in \cite{Vizoso1990}, Wilcox Solar Observatory from 1976 to 2009 in \cite{Hoeksema2010}) have shown a north-south asymmetry in the sunspots distribution. The southern hemisphere led by 18 months over cycle 19, and since then the northern hemisphere leads by a year on average \citep{Svalgaard2013}. The maximum delay between the two hemispheres measured so far is 2 years and it is worth noting that no systematic pattern has been found for this polar reversal phase delay in the surface magnetic activity \citep{Temmer2006}. We can note however that a systematic asymmetry can be observed for the average heliospheric current sheet at the Earth's orbit which appears to be systematically shifted southwards since at least cycle 16 (\cite{Mursula2003}). 

A possible explanation of this asymmetry has been proposed : it can be due to the interference between the dipolar and quadrupolar components of the magnetic field can lead in extreme cases to the vanishing of the magnetic field in one hemisphere \citep{Gallet2009}. See also \cite{Tobias1997}, \cite{Ossendrijver2003}, \cite{DeRosa2012}, \cite{Shukuya2017} and references therein for a detailed discussion on dynamo symmetry properties. The antisymmetric family corresponds to odd $\ell$ degrees when projecting the magnetic field onto the spherical harmonics functions $Y_\ell^m$ under the assumption of axisymmetry (with order $m=0$). It is overall dominant in the Sun, which explains the apparent dipolar structure over most of the cycle. However the symmetric family, which corresponds to even $\ell$ degrees (when $m=0$), is not negligible; it reaches on average 25\% of the amplitude of the antisymmetric family, and becomes dominant during polarity reversals. Such asymmetry has also an impact on the wind structure : in \cite{Sokol2015}, the reconstruction of the solar wind using Interplanetary Scintillations (IPS) observations summed up in \cite{Tokumaru2010}, shows clearly an asymmetry in the latitudinal distribution of the wind.

For the large-scale magnetic field generation, the general theoretical framework is the dynamo theory, especially the interface dynamo \citep{Parker1993} : due to strong shears in the solar differential rotation \citep{Schou1998}, a toroidal field is generated in the tachocline; the poloidal field is regenerated by induction thanks to the turbulent fluid motions in the convection zone, thus sustaining it against ohmic dissipation (see \cite{Miesch2005} for review on the subject). The most realistic models are associated with a flux-transport mechanism and a Babcock-Leighton term to take into account the role of the meridional circulation, linking the surface of the Sun and the tachocline at the poles by redistributing the magnetic field (see \cite{Babcock1961} and \cite{Leighton1969}, but also \cite{Wang1991} for the link with observations and \cite{Dikpati1999} for simulations). 

To simulate the whole Sun, magnetohydrodynamic (or MHD) simulations have been used successfully since the 1970s (for a full review on the subject, see \cite{Brun2017}).

Large Eddy Simulations (LES) are global-scale simulations where  a filter is applied to remove the small scales which are treated as dissipation terms. They have been first computed in \cite{Gilman1983} and in \cite{Glatzmaier1985}, and first adapted for high resolutions in \cite{Brun2004}; more realistic set-ups with high Reynolds numbers can be found in \cite{Hotta2016}. LES models are now able to display large-scale magnetic cycles : for the Sun see \cite{Ghizaru2010}, for young convective stars see \cite{Brown2011}, for exceptional cycles such as grand minima see \cite{Augustson2015} and for generalization to other solar-like stars see \cite{Strugarek2017}.

Mean Field Simulations (MFS) are simulations performed under the assumptions of mean-field theory and axisymmetry. They have the advantage of being computationally cheap compared to the other fully 3D MHD methods. A presentation of kinematic dynamo models can be found in \cite{Roberts1972}, for more focus on the role of rotation see \cite{Stix1976}; for general reviews see \cite{Krause1980} and \cite{Charbonneau2010}. Though they rely on a simplified physical model, MFS reproduce better the magnetic field at the surface of the Sun for now. New models are even able to produce sunspots \citep{Kumar2018}.

For the solar wind, the current framework has been initiated by the work of \cite{Parker1958} : the observed gap of pressure and temperature between the solar corona and the interstellar medium led to the emerging idea of a dynamic corona, expanding to supersonic speed. This hydrodynamical approach was then expanded to take into account the magnetic field, which yields better computations of the angular momentum loss (\cite{Schatzman1962}, \cite{Weber1967}). It was also realized that, to explain the fast distribution of the wind, we certainly needed to take into account subtle magnetic effects and to adopt a multi-fluid approach \citep{Hollweg2002}. These 1D analytical equatorial models were then expanded to 2D \citep{Sakurai1985}, leveraging the conservation of various physical quantities along the poloidal streamlines. There are still a lot of mechanisms that are not fully understood and yet to be modeled. For instance, the coronal heating is still an open problem : we do not know the exact mechanism, although there are very promising studies around magnetic reconnection \citep{Parker1988} and MHD turbulence \citep{Cranmer2007}. The open flux problem is also a rising challenge, with the question of observational magnetic maps underestimating the open magnetic flux of the Sun \citep{Linker2017}.  

One way to reproduce the wind dynamics is via compressible MHD simulations \citep{Keppens1999} and there are basically two approaches : study of one or several flux tubes starting from the chromosphere to focus on the energy transfer between the surface and the corona in one \citep{Suzuki2005} or either in two dimensions \citep{Matsumoto2012}, or more global simulations including the whole star either in 2.5D in \cite{Matt2012}, \cite{Toth2012}, \cite{Reville2015a} or in 3D in \cite{Riley2015} and \cite{Reville2017}. 

Though connected in the real Sun, the solar interior and atmosphere are very difficult to couple numerically. Different plasma parameters, the wide range of time and length scales, the variety of physical processes involved and the stiffness of the MHD equations all together make the modeling of the complete Sun an obvious numerical challenge. Remarkable attempts to deal with this problem are numerical studies made in small Cartesian domains (few tens of Megameters) including the different photospheric layers (\cite{Vogler2005}, \cite{Martinez2008} for the addition of conduction, \cite{Rempel2009} for focus on sunspots among others ; see also review by \cite{Wedemeyer2009}). 

In a previous study, published in \cite{Pinto2011}, the influence of the magnetic field geometry and amplitude on the solar wind has been studied in a quasi-static way in 2.5D. A first numerical code named STELEM (STellar ELEMents, cf. \cite{Jouve2007}) computes an 11-year magnetic cycle using a kinematic mean-field $\alpha-\Omega$ dynamo model. The latter is used as a bottom boundary condition for a second numerical code (DIP, cf. \cite{Grappin2010}) which computes a sequence of relaxed steady states of an isothermal wind. Quantities such as the wind speed distribution, the mass loss and the angular momentum loss were thus computed over an activity cycle, showing considerable variations over time. 

In this paper, we use a similar approach but with a more realistic dynamo model as described in \cite{Jouve2007} and emphasizing north-south asymmetry as in \cite{DeRosa2012} (see their Appendix A). We will first present the numerical codes and the physical ingredients in our models in Section \ref{sec:setup}, then we will present our results as followed : we will begin with a description of the magnetic dynamo cycle in Section \ref{sec:cycle} (time-latitude diagram, topology of the field and its evolution), then we will focus on the variation of the solar wind speed and its spatial distribution in Section \ref{sec:wind_speed} and finally we will describe the evolution of global quantities such as mass and angular momentum loss over the cycle under the influence of asymmetry in Section \ref{sec:global}. Discussion and conclusion are made in section \ref{sec:discussion}.
 
\section{Numerical set-up}\label{sec:setup}

We used two different MHD codes for our 2.5D axisymmetric simulations : STELEM for the dynamo field generated inside the star with a flux-transport model (Emonet \& Charbonneau 1998, private communication; \cite{Jouve2007}), and PLUTO for the solar corona and the associated wind \citep{Mignone2007}. The coupling between the two codes is made through the magnetic field properties : the dynamo field is matching a potential field at the surface at the star, which is then used as a bottom boundary condition for the corona and wind, and initial background field over the whole computational domain. 

In Section \ref{sec:stelem} we describe physical properties and the numerical methods used in the STELEM code; the same is done for the PLUTO code in Section \ref{sec:pluto}. Finally we give further details about the coupling process in Section \ref{sec:coupling}.

\subsection{Dynamo simulations with STELEM}\label{sec:stelem}

The simulations performed are the same as the one described in \cite{DeRosa2012} in their appendix.

Working in spherical coordinates $(r,\theta,\phi)$ and under the assumption of axisymmetry, we perform a poloidal-toroidal decomposition and write the mean magnetic and velocity field (respectively $\boldsymbol{B}$ and $\boldsymbol{U}$) as follows :
\begin{equation}
\boldsymbol{B}(r,\theta,t) = \nabla\times(A_{\phi}(r,\theta,t)\boldsymbol{e}_{\phi}) + B_{\phi}(r,\theta,t)\boldsymbol{e}_{\phi},
\label{eq:poltor_B}
\end{equation}
\begin{equation}
\boldsymbol{U}(r,\theta) = \boldsymbol{u}_p(r,\theta) + r\mathrm{sin}\theta\Omega(r,\theta)\boldsymbol{e}_{\phi}.
\label{eq:poltor_U}
\end{equation}

$\boldsymbol{B}$ is decomposed with the poloidal stream function $A_{\phi}$ and toroidal field $B_{\phi}$. The velocity field is time-independent and prescribed by profiles of the meridional circulation $\boldsymbol{u}_p$ and differential rotation $\Omega$.

We rewrite the induction equation in the framework of mean-field theory in terms of $A_{\phi}$ and $B_{\phi}$ and we introduce a poloidal term $S$ based on the Babcock-Leighton mechanism. We finally normalize the equations by using the solar radius $R_\odot$ as length scale and the diffusion time $R_\odot^2/\eta_t$ as the time scale, and obtain the following two coupled partial differential equations :
\begin{equation}
\frac{\partial A_{\phi}}{\partial t} = \frac{\eta}{\eta_t}\left(\nabla^2-\frac{1}{\varpi^2}\right)A_{\phi} - R_e\frac{\boldsymbol{u}_p}{\varpi}\cdot\nabla(\varpi A_{\phi})+C_sS,
\label{eq:aphi}
\end{equation}
\begin{align}
\frac{\partial B_{\phi}}{\partial t} = & \frac{\eta}{\eta_t}\left(\nabla^2-\frac{1}{\varpi^2}\right)B_{\phi} + \frac{1}{\varpi}\frac{\partial(\varpi B_{\phi})}{\partial r}\frac{\partial(\eta/\eta_t)}{\partial r} - R_e\varpi\boldsymbol{u}_p\cdot\nabla\left(\frac{B_{\phi}}{\varpi}\right) \nonumber \\& - R_eB_{\phi}\nabla\cdot\boldsymbol{u}_p + C_{\Omega}\varpi[\nabla\times(A_{\phi}\hat{\boldsymbol{e}_{\phi}})]\cdot\nabla\Omega,
\label{eq:bphi}
\end{align}
where $\eta$ is the effective magnetic diffusivity, $\eta_t$ is the turbulent diffusivity in the convection zone and $\varpi = r\mathrm{sin}\theta$. These equations are controlled by three dimensionless parameters : $C_{\Omega}=\Omega_0R_\odot^2/\eta_t$ which characterizes the shear by differential rotation (i.e. the omega effect) ; $C_s=s_0R_\odot/\eta_t$ which characterizes the Babcock-Leighton source term ; $R_e=u_0R_\odot/\eta_t$ (the Reynolds number) which characterizes the intensity of the meridional circulation. $\Omega_0$, $s_0$ and $u_0$ are respectively the rotation rate, the typical amplitude of the surface source term and the amplitude of the meridional flow. In this study, we have $C_{\Omega}=1.4 \ 10^5$, $C_s = 30$ and $R_e = 1.20 \ 10^3$, which corresponds to the parameters in \cite{DeRosa2012}. The rotation rate thus correspond to the one measured by helioseismology.

For simplicity, we assume that the meridional circulation is equatorially symmetric, having one large single cell in each hemisphere. Flows are directed poleward at the surface and equatorward at depths, vanishing at the bottom radial boundary. The equatorward flow penetrates slightly beneath the tachocline \citep{Dikpati1999}. 

The rotation profile is inspired by solar angular velocity profile deduced from helioseismic inversions \citep{Thompson2003}. We assume solid-body rotation below $r=0.66R_\odot$ and a differential rotation above :
\begin{equation}
\Omega(r,\theta) = \Omega_c+\frac{1}{2}\left[1+\mathrm{erf}\left(\frac{2(r-r_c)}{d_1}\right)\right](\Omega_{eq}+a_2\mathrm{cos}^2\theta+a_4\mathrm{cos}^4\theta-\Omega_c),
\label{eq:diffrot}
\end{equation}
with the parameters : $\Omega_{eq}=1$, $\Omega_c=0.93944$, $r_c=0.7R_\odot$, $d_1=0.05R_\odot$, $a_2=-0.136076$ and $a_4=-0.145713$.

We assume different diffusivities in the envelope and in the stable interior, smoothly matching the two :
\begin{equation}
\eta(r) = \eta_c+\frac{(\eta_t-\eta_c)}{2}\left[1+\mathrm{erf}\left(\frac{r-r_c}{d}\right)\right],
\end{equation}
with the parameters : $\eta_c=10^9 \ \mathrm{cm}^2\mathrm{s}^{-1}$, $\eta_t=10^{11} \ \mathrm{cm}^2\mathrm{s}^{-1}$ and $d=0.03R_\odot$.

In Babcock-Leighton flux-transport dynamo models, the poloidal field owes its origin to the tilt of magnetic loops emerging at the solar surface. Thus the source has to be confined to a thin layer just below the surface and, since the process is fundamentally non-local, the source term depends on the variation of $B_{\phi}$ at the base of the convection zone. To create asymmetry between the two hemispheres we introduce a modified source term modulated by the asymmetry parameter $\epsilon$:
\begin{align}
S(r,\theta,B_{\phi}) = &\frac{1}{2}\left[1+\mathrm{erf}\left(\frac{r-r_2}{d_2}\right)\right]\left[1-\mathrm{erf}\left(\frac{r-R_\odot}{d_2}\right)\right]\left[1+\left(\frac{B_{\phi}(r_c,\theta,t)}{B_0}^2\right)\right]^{-1} \label{eq:bl} \\
&(\mathrm{cos}\theta+\epsilon\mathrm{sin}\theta)\mathrm{sin}^3\theta B_{\phi}(r_c,\theta,t), \nonumber
\end{align}
with the parameters : $r_2=0.95R_\odot$, $d_2=0.01R_\odot$, $B_0=10^5$ and $\epsilon = 10^{-3}$.

The STELEM code is used to solve equations \eqref{eq:aphi} and \eqref{eq:bphi} : it uses a finite element method in space and a third-order scheme in time. The temporal scheme used is adapted from \cite{Spalart1991} and is similar to a Runge-Kutta 3 method. The STELEM code has been thoroughly tested and validated via an international mean-field dynamo benchmark involving eight different codes \citep{Jouve2008}. 

The numerical domain is an annular meridional cut of the Sun with the colatitude $\theta \in [0,\pi]$ and the radius $r\in [0.6,1]R_\odot$ (ie. from slightly below the tachocline located at $r\approx 0.7R_\odot$ up to the solar surface). We use a uniform grid with 64 points in radius and cosine of the latitude. At the latitudinal boundaries ($\theta=0$ and $\theta=\upi$) and at the bottom radial boundary ($r=0.6R_\odot$), $A_{\phi}$ and $B_{\phi}$ are set to 0. At the upper radial boundary ($r=R_\odot$), the solution is matched to an external potential field. Usual initial conditions involve setting a confined dipole field configuration, i.e. $A_{\phi}$ is set to $\mathrm{sin}\theta/r^2$ in the convection zone and to 0 below the tachocline ; the toroidal field is initialized to 0 everywhere. 

\subsection{Wind simulations with PLUTO}\label{sec:pluto}

This set-up is inspired by the one described in \cite{Reville2015a}, but adapted to spherical coordinates.

We solve the set of the conservative ideal MHD equations composed of the continuity equation for the density $\rho$, the momentum equation for the velocity field $\boldsymbol{u}$ with its momentum written $\boldsymbol{m}=\rho\boldsymbol{u}$, the energy equation which is noted $E$ and the induction equation for the magnetic field $\boldsymbol{B}$:
\begin{equation}
\frac{\partial}{\partial t}\rho+\nabla\cdot\rho\boldsymbol{u}=0,
\end{equation} 
\begin{equation}
\frac{\partial}{\partial t}\boldsymbol{m}+\nabla\cdot(\boldsymbol{mv}-\boldsymbol{BB}+\boldsymbol{I}p) = \rho\boldsymbol{a},
\end{equation}
\begin{equation}
\frac{\partial}{\partial t}E + \nabla\cdot((E+p)\boldsymbol{u}-\boldsymbol{B}(\boldsymbol{u}\cdot\boldsymbol{B})) = \boldsymbol{m}\cdot\boldsymbol{a},
\end{equation}
\begin{equation}
\frac{\partial}{\partial t}\boldsymbol{B}+\nabla\cdot(\boldsymbol{uB}-\boldsymbol{Bu})=0,
\end{equation}
where $p$ is the total pressure (thermal and magnetic), $\boldsymbol{I}$ is the identity matrix and $\boldsymbol{a}$ is a source term (gravitational acceleration in our case). We use the ideal equation of state :
\begin{equation}
\rho\varepsilon = p_{th}/(\gamma -1),
\end{equation}
where $p_{th}$ is the thermal pressure, $\varepsilon$ is the internal energy per mass and $\gamma$ is the adiabatic exponent. This gives for the energy : $E = \rho\varepsilon+\boldsymbol{m}^2/(2\rho)+\boldsymbol{B}^2/2$.

PLUTO solves normalized equations, using three variables to set all the others: length, density and speed. If we note with $*$ the parameters related to the star and with $0$ the parameters related to the normalization, we have $R_*/R_0=1$, $\rho_*/\rho_0=1$ and $u_{kep}/U_0=\sqrt{GM_*/R_*}/U_0=1$, where $u_{kep}$ is the Keplerian speed at the stellar surface and $G$ the gravitational constant. By choosing the physical values of $R_0$, $\rho_0$ and $U_0$, one can deduce all of the other values given by the code in physical units. In our set-up, we choose $R_0=R_\odot=6.96 \ 10^{10}$ cm, $\rho_0=\rho_\odot=6.68 \ 10^{-16} \ \mathrm{g/cm}^3$ (which corresponds to the density in the solar corona above 2.5 km, cf. \cite{Vernazza1981}) and $U_0=u_{kep,\odot}=4.37 \ 10^2$ km/s. Our wind simulations are then controlled by three parameters : the adiabatic exponent $\gamma$ for the polytropic wind, the rotation of the star normalized by the escape velocity $u_{rot}/u_{esc}$ and the speed of sound normalized also by the escape velocity $c_s/u_{esc}$. Note that the escape velocity is defined as $u_{esc} = \sqrt{2}u_{kep} = \sqrt{2GM_*/R_*}$. For the rotation speed, we take the solar value, which gives $u_{rot}/u_{esc} = 2.93 \ 10^{-3}$. We choose to fix $c_s/u_{esc}=0.243$, which corresponds to a $1.3 \ 10^6$ K hot corona for solar parameters and $\gamma=1.05$. This choice of $\gamma$ is dictated by the need to maintain an almost constant temperature as the wind expands, which is what is observed in the solar wind. Hence, choosing $\gamma \neq 5/3$ is a simplified way of taking into account heating, which is not modeled here. 

We assume axisymmetry and use the spherical coordinates $(r,\theta,\phi)$. Since PLUTO is a multi-physics and multi-solver code, we choose a finite-volume method using an approximate Riemann Solver (here the HLL solver, cf. \cite{Einfeldt1988}). PLUTO uses a reconstruct-solve-average approach using a set of primitive variables $(\rho,\boldsymbol{u},p,\boldsymbol{B})$ to solve the Riemann problem corresponding to the previous set of equations. 

The numerical domain is an annular meridional cut with the colatitude $\theta \in [0,\upi]$ and the radius $r \in [1,20]R_\odot$. We use an uniform grid in latitude with 512 points, and a stretched grid in radius with 512 points; the grid spacing is geometrically increasing from $\Delta r/R_*=0.001$ at the surface of the star to $\Delta r/R_*=0.01$ at the outer boundary. At the latitudinal boundaries ($\theta=0$ and $\theta=\upi$), we set axisymmetric boundary conditions. At the top radial boundary ($r=20 R_*$), we set an outflow boundary condition which corresponds to $\partial/\partial r=0$ for all variables, except for the radial magnetic field where we enforce $\partial(r^2B_r)/\partial r=0$. Because the wind has opened the field lines and under the assumption of axisymmetry, this ensures the divergence-free property of the field. At the bottom radial boundary ($r=R_*$), we set a condition similar to the one described in \cite{Zanni2009} to be as close as possible to a perfect rotating conductor. We also implement the same differential rotation as described in \eqref{eq:diffrot}. We initialize the velocity field with a polytropic wind solution and the magnetic field with a potential extrapolation of the field produced by STELEM at a given time. Please note that there is a splitting in PLUTO between the background field (which is curl-free and provided by the dynamo model) and the deviation field (which is a perturbation of the background field and carries the magnetic energy). Please note also that, to enforce the divergence-free property of the field, we use a hyperbolic divergence cleaning, which means that the induction equation is coupled to a generalized Lagrange multiplier in order to compensate the deviations from a divergence-free field \citep{Dedner2002}.

\subsection{Coupling method}\label{sec:coupling}

To couple the two codes described above, we use the projection of the surface magnetic field produced by STELEM on spherical harmonics, which PLUTO can read. To proceed so, we select a given time $t_0$, then we project the surface radial magnetic field at that time $B_r(t_0,r=R_\odot,\theta,\phi)$ on spherical harmonics $Y_\ell^0(\theta)$ up to a degree $\ell_{max}$ :
\begin{equation}
B_r(r=R_{\odot},\theta,t_0) = \sum_{\ell=1}^{\ell_{max}}\alpha_{\ell,0}(r=R_{\odot})Y_\ell^0(\theta).
\end{equation} 

Note that there is no dependency in $\phi$ due to axisymmetry ($m=0$). In this study, we choose $\ell_{max}=40$, as it provides the best compromise between accuracy and costs for the reconstruction. Then inside PLUTO, we reconstruct the field by reading these coefficients and combining them with spherical harmonics, and finally we extrapolate the field inside the whole wind computation domain. For the extrapolation, we chose a Potential-Field Source Surface (PFSS) method with a source surface radius $R_{ss}$ equal to $15R_\odot$. For more information about the PFSS method, for original computation see \cite{Schatten1969} and \cite{Altschuler1969}, for more explicit calculation see \cite{Schrijver2003}. We obtain:
\begin{equation} 
\alpha_{\ell,0}(r) = \alpha_{\ell,0}(r=R_{\odot})\frac{\ell(R_{\odot}/R_{ss})^{2\ell+1}(r/R_{\odot})^{\ell-1}+(\ell+1)(r/R_{\odot})^{-(\ell+2)}}{\ell(R_{\odot}/R_{ss})^{2\ell+1}+(\ell+1)},
\end{equation}
\begin{equation}
\beta_{\ell,0}(r) = (\ell+1)\alpha_{\ell,0}(r=R_{\odot})\frac{(R_{\odot}/R_{ss})^{2\ell+1}(r/R_{\odot})^{\ell-1}-(r/R_{\odot})^{-(\ell+2)}}{\ell(R_{\odot}/R_{ss})^{2\ell+1}+(\ell+1)}.
\end{equation}

In \cite{Reville2015b}, it was explained that the fiducial value of 2.5 $R_\odot$ usually found in the literature for $R_{ss}$ was an approximation, and that you can define an optimal source surface radius that allows you to recover the best value for the open flux in your simulation. This optimal value increases with the magnetic field strength and decreases with high order magnetic topology and rotation rate. It was suggested that a different value of $R_{ss}$ should thus be used at different times over the cycle, but we did not investigate such parametrization as the PFSS is only used to initialize our wind model. Our value is an estimate of the larger value needed in our simulations, which correspond to a strong dipole at low rotation rate. We recall also that the PFSS extrapolation is only used to initialize the simulation, and the final configuration does not depend much on the initial extrapolation.

This yields the following field reconstruction :
\begin{equation}
\mathbf{B}^{PLUTO} =
\left\{
\begin{aligned}
B_r(r,\theta,\phi) & = \sum_{\ell=1}^{\ell_{max}}\alpha_{\ell,0}(r)Y_\ell^0(\theta) \\
B_{\theta}(r,\theta,\phi) & = \sum_{\ell=1}^{\ell_{max}}\beta_{\ell,0}(r)\sqrt{\frac{2\ell+1}{4\pi}}\frac{1}{\ell+1}\partial_{\theta}P_\ell^0(\theta) \\
B_{\phi}(r,\theta,\phi) & = 0.0.
\end{aligned}
\right. 
\label{eq:pfss}
\end{equation}
The magnetic field provided by STELEM is dimensionless (cf. \eqref{eq:aphi} and \eqref{eq:bphi}). To recover the proper physical units, we calibrate the radial magnetic field amplitude so that the mass loss rate is as close as possible to estimations of the global mass loss rate deduced from Ulysses data \citep{McComas2008} ; this yields values between $2.3 \ 10^{-14} \ M_\odot/yr$ and $3.1 \ 10^{-14} \ M_\odot/yr$ \citep{Reville2017}. The magnetic field is then normalized by the PLUTO units described in section \ref{sec:pluto}.

From our dynamo run, we obtain thus a times series of 54 snapshots ${B_r(t_i)}$, sampled over an 11-year cycle with a time step of about 2 months. For each input magnetic field, we let the wind relax and reach a steady state ; this takes around 500 characteristic times (defined as $t_P = R_*/u_{kep}=R_*^{3/2}/\sqrt{GM_*}$), which translates to 9 days using solar units. The result is then a sequence of steady-state solutions, providing general properties of the coronal magnetic field and the wind between solar minimum and maximum of activity, but without taking into account the back-reaction of the wind on the dynamo. All simulations are performed from scratch, so there is no dependence for any state of relaxation on the previous states.

\section{Description of the cycle}\label{sec:cycle}

\subsection{Properties of the dynamo field}\label{sec:dynamo}

\begin{figure}
\centering
\begin{subfigure}[t]{0.49\textwidth}
  \centering
  \includegraphics[width=\textwidth]{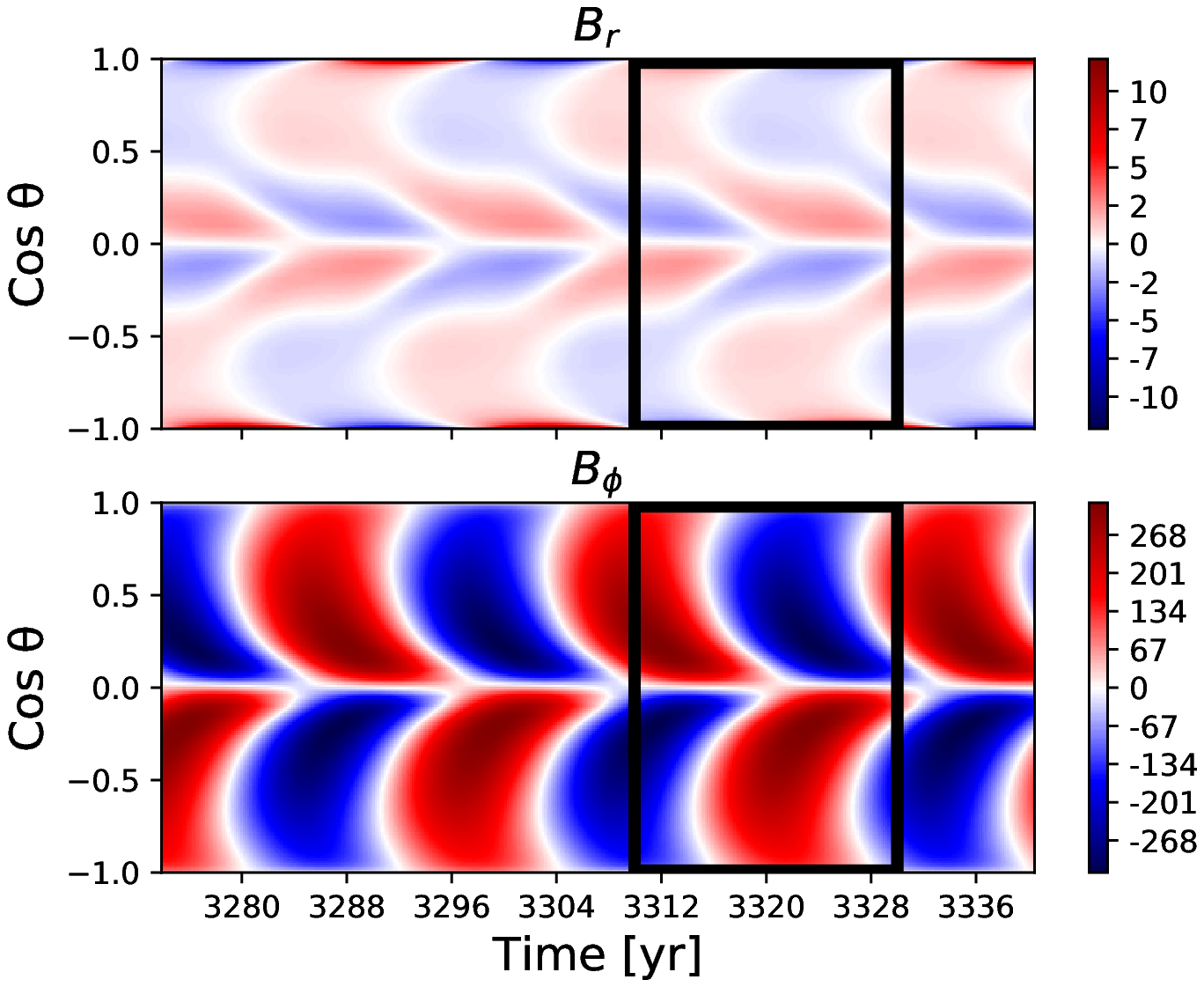} 
\end{subfigure}
\begin{subfigure}[t]{0.49\textwidth}
  \centering
  \includegraphics[width=\textwidth]{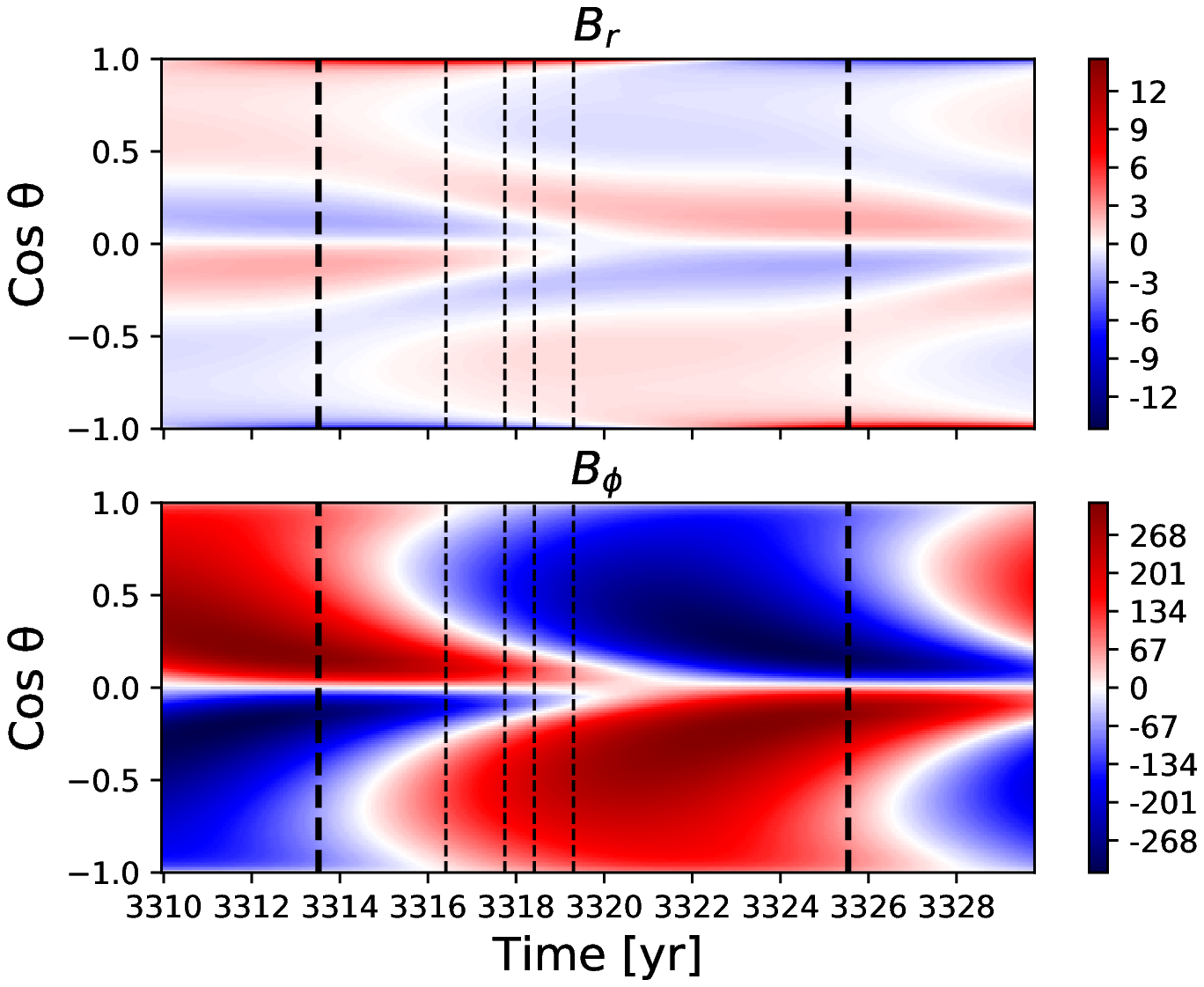} 
\end{subfigure}
\caption{Time-latitude diagrams of the radial surface and the azimuthal tachocline magnetic fields for the dynamo model produced by STELEM. The time is shown in years and the vertical axis shows the cosine of the angle associated to the latitude. The left panel shows the general aspect of the magnetic field over 3 cycles. The right panel is a zoom on the specific cycle we studied, with the black dashed lines indicating remarkable times of the simulation.}
\label{fig:diagpap}
\end{figure}

First we will begin by presenting the main features of the solar dynamo solution generated by STELEM. 

\begin{figure}
\centering
\includegraphics[width=10cm]{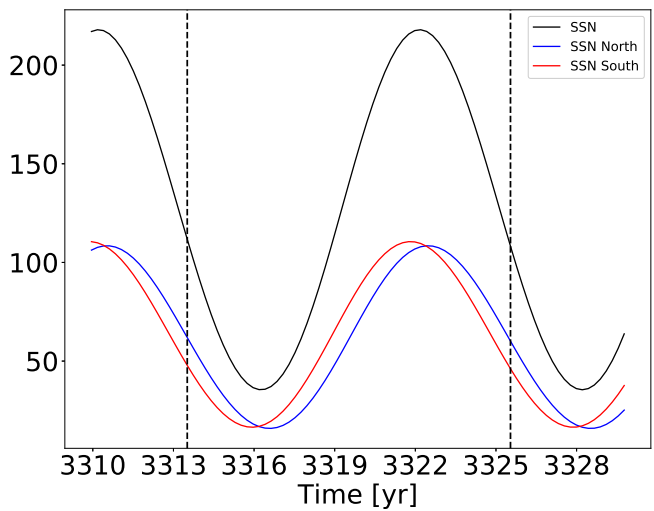}
\caption{Evolution of the proxy for the sunspot number (SSN) over time. The SSN for the northern (southern) hemisphere is in blue (red), and the total number is in black. The black dashed lines indicate the minima of the studied cycle as defined by topology.}
\label{fig:ssn}
\end{figure}

The left panel of Figure \ref{fig:diagpap} displays the time-latitude diagrams for the radial magnetic field at the surface of the star and the toroidal magnetic field at the base of the convective zone over several dynamo cycles. The cycle period is approximately 12.0 years, which is close to the observational mean Sun value of approximately 11 years \citep{Clette2012}. The asymmetry of the model results in a delay of 9 months of the magnetic configuration between the two hemispheres (for example the southern one reverses first, like in cycle 18 of the Sun). This is in qualitative agreement with the current observed delay of 1 year \citep{Temmer2006}. This diagram for the surface radial magnetic field is typical of a flux-transport dynamo model, and exhibits features similar to the solar field : at each instant of the cycle there are two branches, one equatorward and the other one poleward. The right panel of Figure \ref{fig:diagpap} shows more precisely which cycle we decided to study, with black dashed lines indicating some remarkable times at which we computed the associated wind solution with PLUTO. These times are 0.0, 2.9, 4.2, 4.9, 5.8 and 12.0 years after the cycle minimum. We will explain in section \ref{sec:corona} why we chose to show these specific moments among the 54 we computed. The most right and most left lines correspond to the cycle minima.

Note that there are several ways to define a cycle minimum in our numerical simulations when comparing with real sunspot time series. In Figure \ref{fig:diagpap}, we fixed the minimum as the time of the cycle when the magnetic field configuration is more dipolar, which means the time when the ratio of the dipole energy over the quadrupole energy is maximum. Likewise, the maximum of the cycle is defined as the maximum ratio of energy between the quadrupole and dipole. Another way to define a minimum of activity is to say that it is the time when there are the lowest number of sunspots on the solar surface. Since our model does not generate sunspots, we use a proxy to determine an equivalent sunspot number (or SSN) based on the generation of toroidal magnetic field at the tachocline, given by:
\begin{equation}
SSN_{proxy} = SSN_0 \int_{\theta=0}^{\theta=\pi}B_\phi(r_{c}, \theta, t)^2r_{c}^2sin\theta d\theta,
\end{equation}
where $SSN_0$ is a scaling constant to find values for our SSN proxy of the same order of magnitude as the Sun \citep{Hung2017}. 

Separating the SSN from the northern and southern hemisphere in Figure \ref{fig:ssn} allows us to see clearly the north-south asymmetry. For the Sun, these two definitions of minimum or maximum of the solar cycle give the same time. However, we can notice that in our model the minimum of sunspot activity is delayed by 2.5 years compared to our minimum of quadrupolar energy (cf. Figure \ref{fig:ssn}). This is because we did not fine-tune the model to match an exact solar cycle, what we seek foremost is to understand the link between the dynamo field and the wind on a fundamental level. However it allows us to isolate the effects of those different contributions and understand more precisely the underlying physics.

\begin{figure}
\centering
\includegraphics[width=10cm]{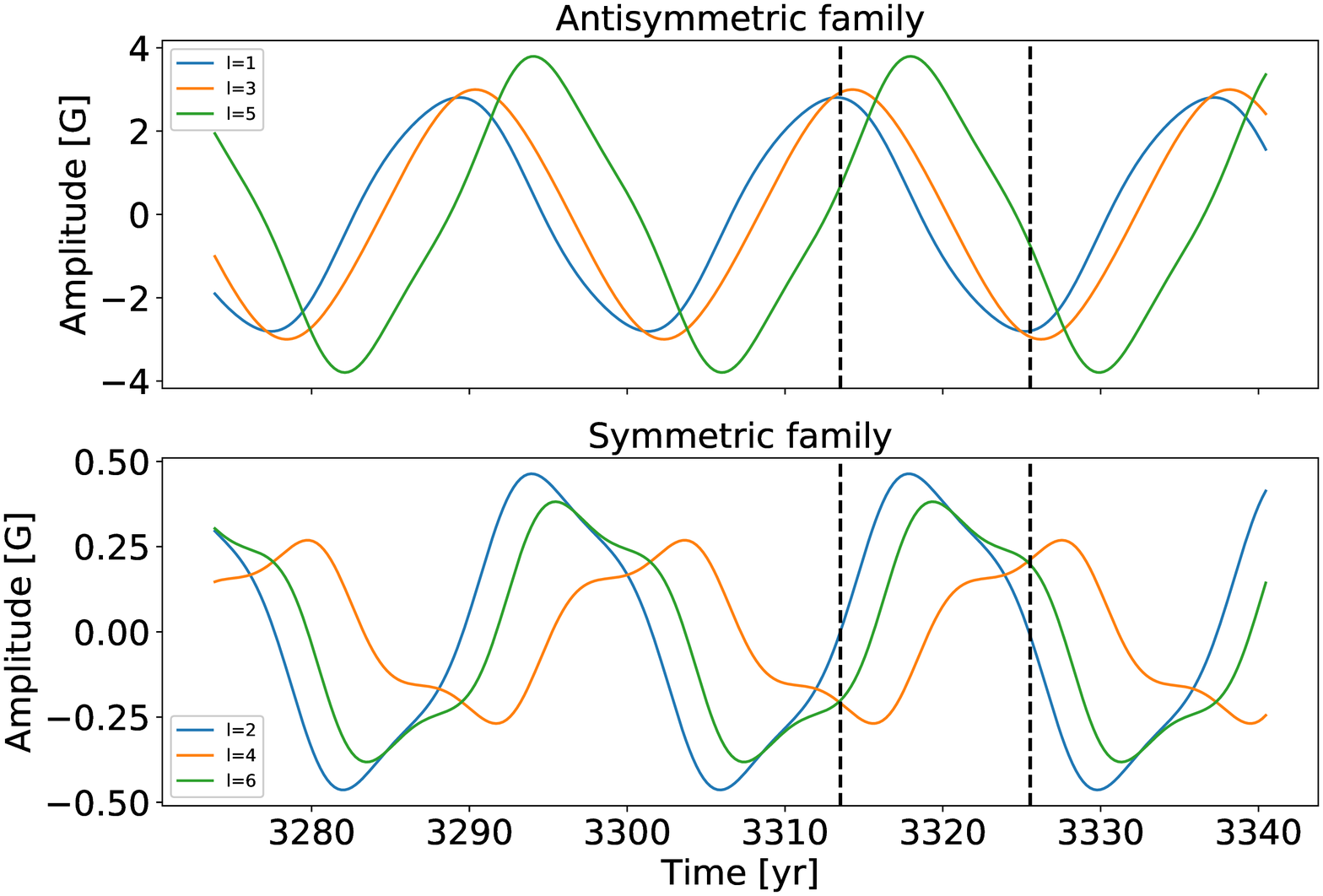}
\caption{Time evolution of the coefficients of the surface magnetic field projection on spherical harmonics over 2.5 dynamo cycles, grouped as equatorially symmetric and antisymmetric families. Only the $m=0$ modes are displayed due to the assumption of axisymmetry.}
\label{fig:coeff_br}
\end{figure}

Another impact of the introduced asymmetry is the ability to couple the equatorially symmetric and antisymmetric family modes for the magnetic field. To demonstrate this point, we display in Figure \ref{fig:coeff_br} the time evolution of the coefficients of the projection of the surface radial magnetic field on the spherical harmonics. They are gathered as equatorially symmetric and antisymmetric families, which correspond to the even and odd degrees $\ell$ for the projection on the spherical harmonics when considering only $m=0$ modes. First thing we can notice is that the amplitude of the antisymmetric family modes is similar to the one of the Sun (between -4 and 4G, as shown in \cite{DeRosa2012}). The $l=3$ component is stronger than what is observed in the Sun : this seems to be induced by the flux-transport model, which tends to accumulate magnetic flux at the poles, thus favoring higher $l$ modes than the dipole and octupole. Despite the fact that the simulation was initialized with a dipole field, we can see that the symmetric family modes develop in a significant way, reaching on average about 10\% of the antisymmetric family mean amplitude. This is different from simple 2.5D mean-field dynamo models : usually the symmetric family modes are unable to develop when the model is initialized with a dipole and is symmetric, whereas in the Sun the quadrupole amplitude is measured to be around 25\% that of the dipole most of the time. We can also note that the amplitude of the symmetric family modes (between -0.5 and 0.5G) is close to what has been observed since cycle 21 (see for instance \cite{DeRosa2012}). 

\begin{figure}
\centering
\begin{subfigure}[t]{0.6\textwidth}
\centering
 \includegraphics[width=\textwidth]{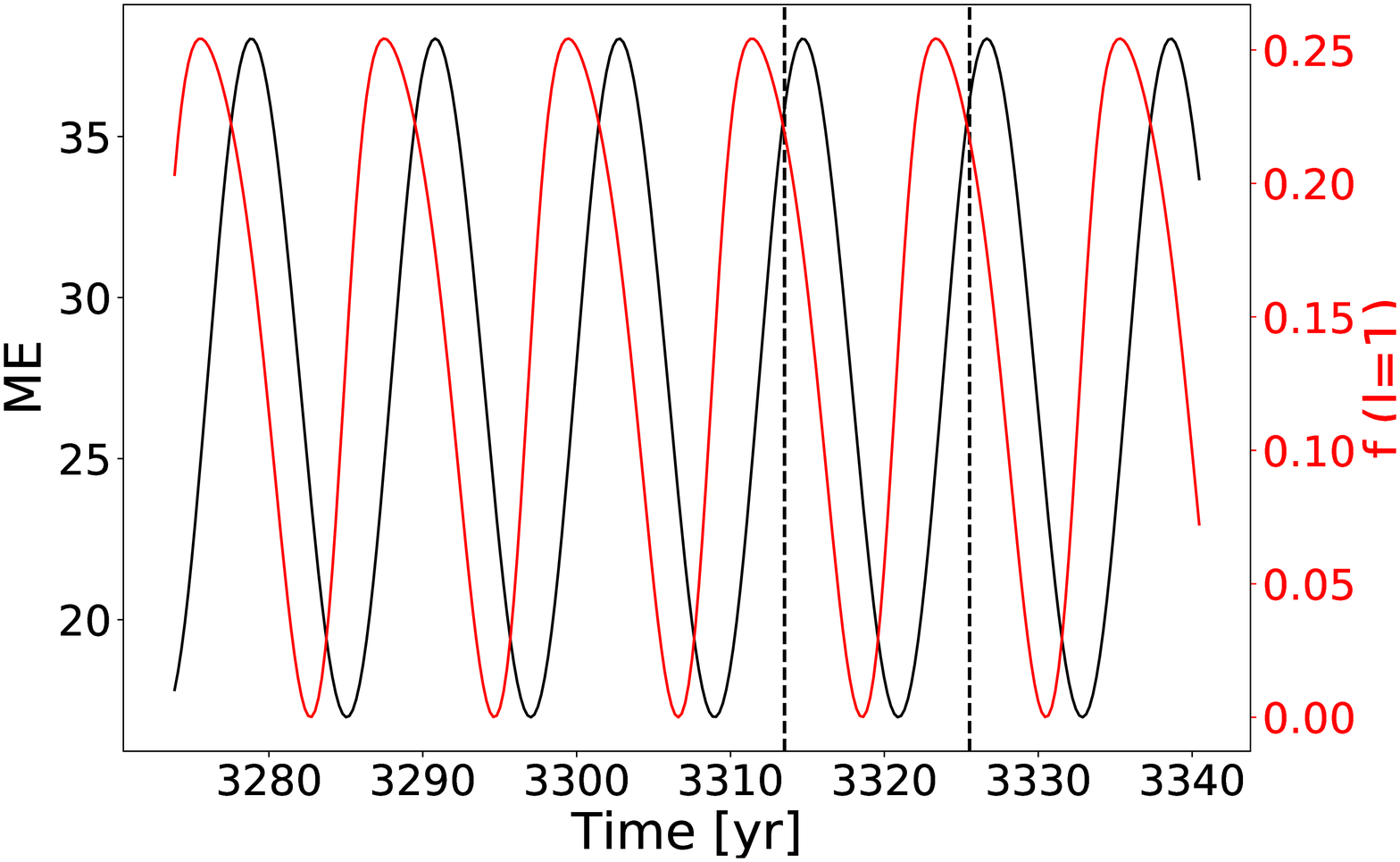}
\end{subfigure}
\begin{subfigure}[t]{0.6\textwidth}
\centering
 \includegraphics[width=\textwidth]{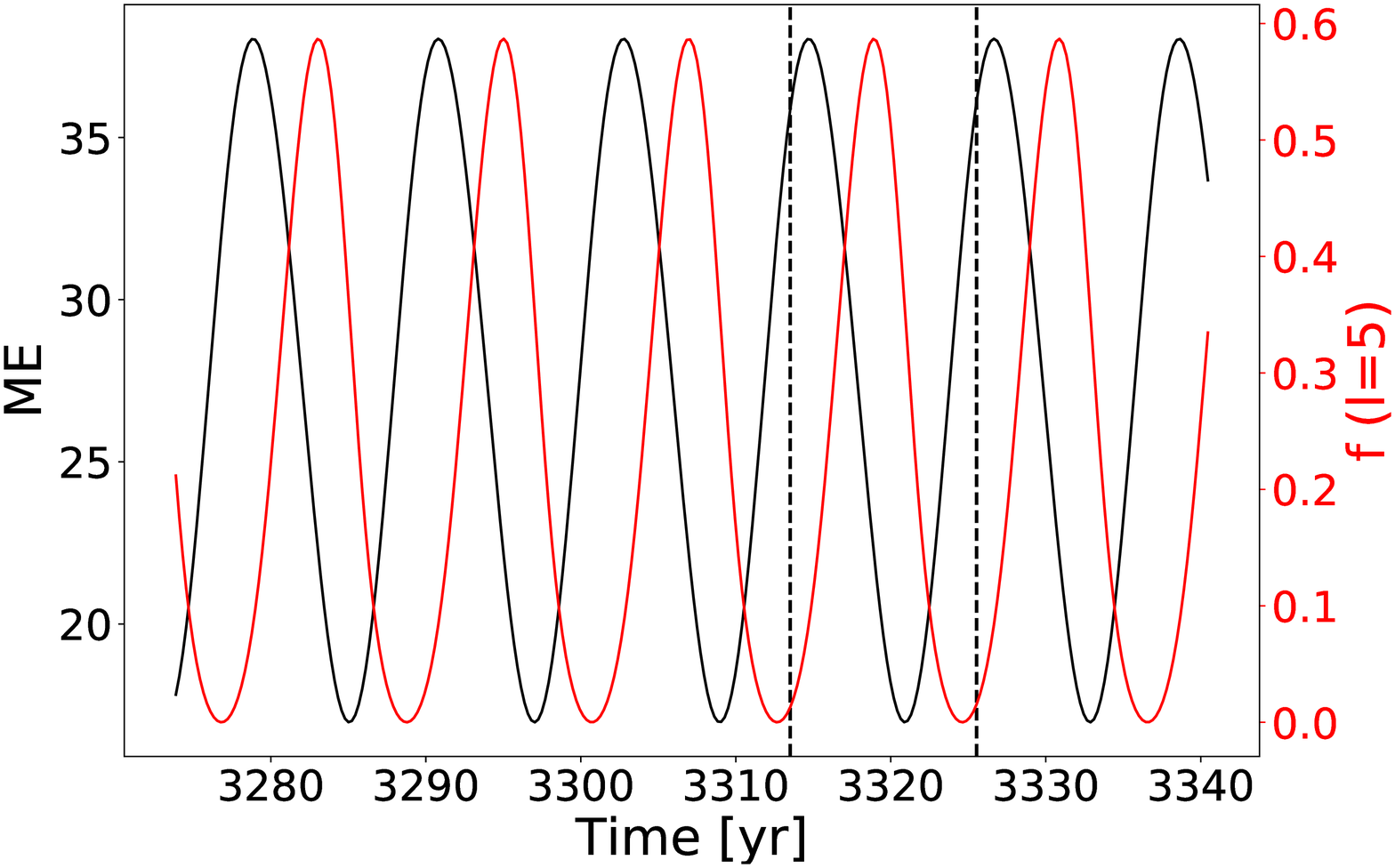}
\end{subfigure}
\caption{Comparison of the evolution of the surface magnetic energy and the energy of 2 mode components over several cycles : $l=1$ for the left panel, $l=5$ for the right panel.}
\label{fig:mag_energy}
\end{figure}

The final property of our dynamo model we wanted to highlight is the time evolution of the total radial surface magnetic energy versus the energy of the dipole component. With our decomposition in spherical harmonics (cf. \eqref{eq:pfss}), we can define the total radial surface magnetic energy as :
\begin{equation}
ME = \sum_{\ell=1}^{\ell_{max}}\alpha_{\ell,0}^2.
\label{eq:ME}
\end{equation}

Then we can define the energy of a specific harmonic component of the radial surface field as :
\begin{equation}
f_{\ell} = \alpha_{\ell,0}^2/ME.
\label{eq:fdip}
\end{equation}

Using data from the Wilcox Observatory it can be shown that $ME$ and $f_{1}$ were anti-correlated from cycle 20 to 23 \citep{Reville2017}, which confirms that the Sun is mostly dipolar during minimum of activity and multipolar near maximum. We plot the time-evolution of these two quantities over 3 cycles in the left panel of Figure \ref{fig:mag_energy}. In our case, $f_{1}$ and $ME$ have a phase delay of one quarter of a period, so not fully correlated nor anti-correlated but in phase quadrature. This is a direct consequence of what we have just highlighted concerning the modes : since the dipole is not here the strongest magnetic mode, it is not the relevant one to study from an energetic point of view. In the right panel of Figure \ref{fig:mag_energy}, we show the same comparison but for the $\ell=5$ mode, and here we have a phase delay of one third of a period, which is closer to anti-correlation. This shows that, although Babcock-Leighton models seem to allow higher modes to reach a bigger amplitude than in the Sun, they capture most of the interplay between topology and energy.

\subsection{Coronal magnetic field}\label{sec:corona}

\begin{figure}
\centering
\includegraphics[width=\textwidth]{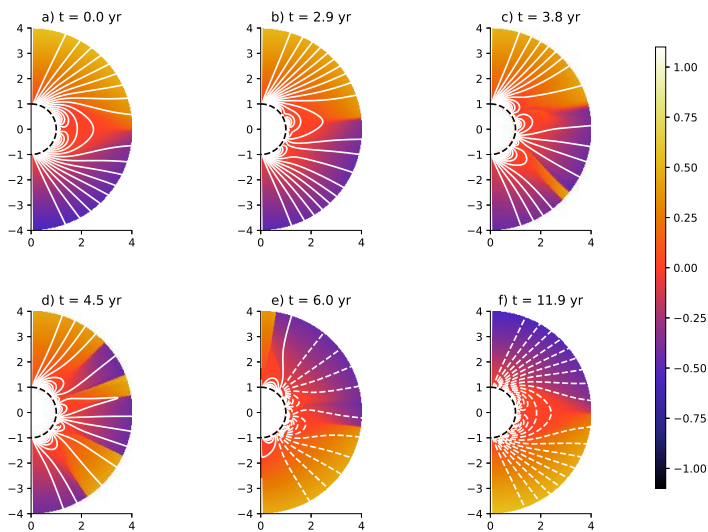}
\caption{Snapshots of the evolution of the coronal magnetic field at different times : if we set $t=0$ years at the first minimum, then we have $t=$0, 2.9, 4.2, 4.9, 5.8 and 12.0 years. The color scale represents the following quantity : $\mathbf{u}\cdot\mathbf{B}/(c_s||\mathbf{B}||)$, which is the solar wind velocity projected on the magnetic field in units of Mach number. White lines correspond to the poloidal magnetic field lines of positive polarity in solid and negative polarity in dashed lines. We represent only the 4 first solar radii.}
\label{fig:snapshots_cycle}
\end{figure}

The time evolution of the coronal magnetic field is displayed in Figure \ref{fig:snapshots_cycle} at remarkable times during the cycle. If we consider the initial minimum of activity at the beginning of our simulation as instant $t=0$, the different panels a, b, c, d, e and f correspond respectively to times $t=$ 0.0, 2.9, 3.8, 4.5, 6.0 and 11.9 years. These times were indicated as black dashed lines in the right panel of Figure \ref{fig:diagpap}. The first 4 $R_\odot$ close to the star are visible for both hemispheres in order to see the effect of asymmetry. The color scale represents the following quantity : $\mathbf{u}\cdot\mathbf{B}/(c_s||\mathbf{B}||)$, which is the solar wind velocity projected on the magnetic field in units of Mach number. Since the wind always flows outwards the star, this quantity allows us to track the changes of polarity of the magnetic field in the open field regions ; with this colortable, yellow corresponds to positive polarity, and dark blue to negative polarity. The transitions between polarities are in red and correspond to current sheets. The poloidal field lines are plotted in white, in solid (dashed) for positive (negative) polarity, corresponding to a positive (negative) potential vector $A_\phi$.

We now analyze the changes of topology in the coronal magnetic field observed over one dynamo cycle and its interaction with the wind. To describe properly the structures, we will differentiate the helmet streamers, which separate coronal holes of opposite magnetic polarities, from the pseudo-streamers, which overlie twin loop arcades and separate holes of the same polarity, as defined in \cite{Wang2007}.

We start at a minimum of activity, go through a maximum and then return to minimum. At minimum of activity (panel a), the magnetic field is dominated by the dipole starting from 2-3 solar radii, but is mostly octupolar at the surface with 3 arcades of closed magnetic field loops. This is consistent with the spectrum analysis displayed in Figure \ref{fig:coeff_br}. We have a central streamer similar to what is observed in the Sun : it extends up to 4 solar radii, where most coronograph pictures show a streamer between 2 and 4 solar radii. The magnetic field is of positive polarity in the northern coronal hole and of negative polarity in the southern coronal hole.

We can then clearly see the reversal of the two hemispheres happening one after the other. In panel b, we can see that the equatorial streamer has been disrupted at high latitudes in the southern hemisphere, leading to the appearance of pseudo-streamers. In the northern hemisphere, the main streamer is still intact and connected transequatorially with the southern hemisphere just under the equator. In panel c, the first coronal hole of opposite polarity opens in the southern hemisphere, creating a streamer at mid-latitude. The equatorial streamer is completely disrupted at this point, pseudo-streamers have been formed as well in the northern hemisphere. In panel d, another coronal hole of opposite polarity forms this time in the northern hemisphere. In the southern hemisphere, the coronal hole is spreading due to the wind opening up the coronal field lines. The magnetic configuration is very close to quadrupolar. In panel e, the southern hemisphere has now reversed, since the overall magnetic field, especially at the poles, is now the opposite of the one in panel a. In the northern hemisphere, the reversal is not complete, the pole is the last region where the coronal hole of opposite polarity has not spread to yet. In panel f, we jump from the middle of the cycle to the end of the cycle: the field has returned to a dipolar configuration, but with the exact opposite polarity compared to panel a, hence completing an activity cycle.

\begin{figure}
\centering
\includegraphics[width=\textwidth]{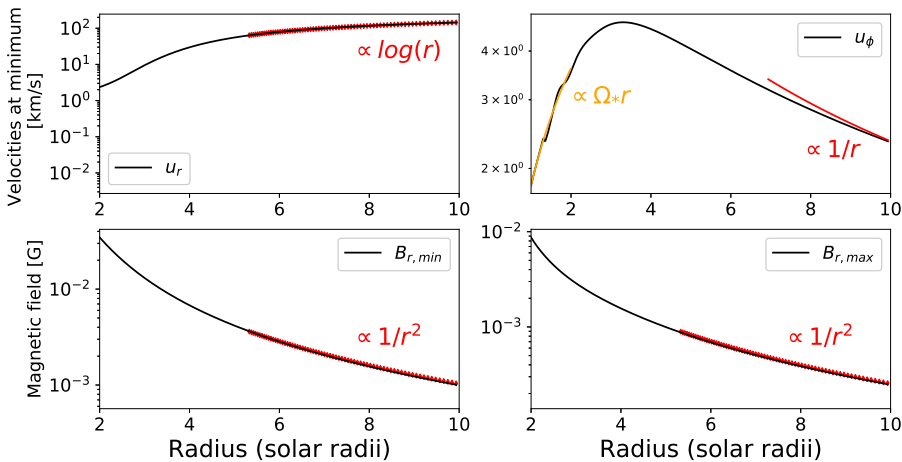}
\caption{Radial cuts at minimum and maximum of activity, for $u_r$, $u_{\phi}$ and $B_r$ between 2 and 10 solar radii.}
\label{fig:radial_cuts}
\end{figure}

We then describe further our model by showing radial cuts of various quantities in the solar corona. We want to check the radial dependency of $u_r$, $u_\phi$ and $B_r$ to see if we recover expected behaviors. In Figure \ref{fig:radial_cuts}, we plot these quantities in the equatorial plane for the velocity and at mid-latitude for the magnetic field to avoid the current sheets, at minimum and maximum of activity. For the velocity profile, we compare it at minimum with predictions of the Weber and Davis wind, as shown in \cite{Keppens1999}. The radial velocity has the profile of an accelerated transonic solution, which at the end is well fitted by a logarithmic function of the radius. The longitudinal velocity is slightly increasing in the first solar radii before slowly decreasing. Its initial slope is equal to the rotation rate of the star, because of the co-rotation of the corona with the star. Further from the star, this becomes less and less true and the surrounding corona rotates slower; we tend to have a $1/r$ dependency. Far from the star, $B_r$ displays an expected dependency of $1/r^2$, typical of a purely radial field stretched by the wind to satisfy the divergence-free property (in spherical coordinates : $\partial(r^2B_r)/\partial r=0$). This is true both at minimum and maximum of activity.

\section{Solar wind speed}\label{sec:wind_speed}

\begin{figure}
\centering
\includegraphics[width=10cm]{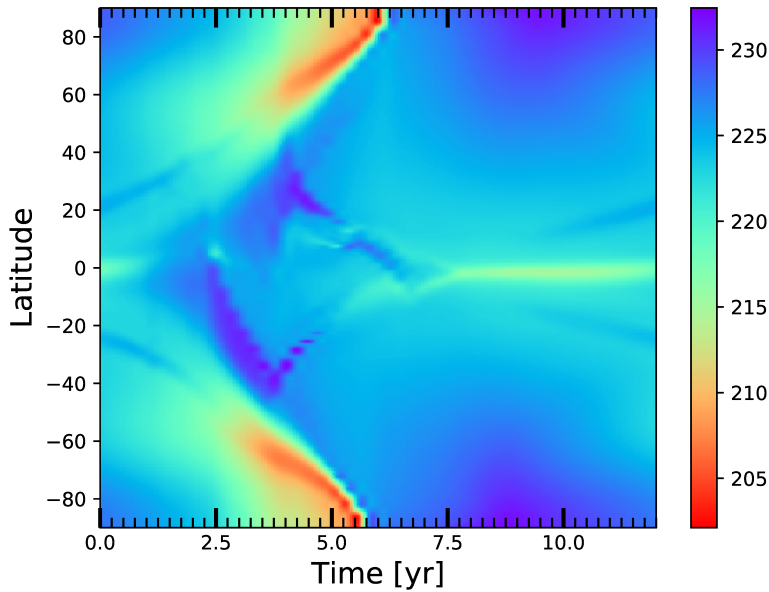}
\caption{Time-latitude diagram of the wind speed over an 11-year cycle in km/s at $r=20R_\odot$.}
\label{fig:diag_wind}
\end{figure}

We will now focus on the variation of the wind speed over a solar cycle. This is one of the most important features because it can be compared with in situ measurements and directly affect planets or satellites. Figure \ref{fig:diag_wind} shows the evolution of the radial wind velocity at $r=20 R_\odot$ in km/s over an 11-year cycle in our simulation. We show only the radial velocity because so far from the star, the wind has become mostly radial, so it is $u_r$ which contains most of the wind total velocity. We chose $r=20 R_\odot$ because it is our domain's outer boundary, which corresponds to almost 0.1 AU; most wind in situ measurements are made at 1 AU, so we have to extrapolate our numerical values to compare them with observational values by using a logarithmic fit. 

First of all, we can see that the amplitude of the wind is different from what is measured : the solar wind flow amplitude is usually between 400 and 800 km/s, while here we reach between 200 and 235 km/s at 0.1 AU, which gives between 420 and 470 km/s at 1 AU. This difference in amplitude is due to our polytropic corona that prevents us from recovering the fast wind, because we have a uniform heating. To recover fast winds numerically, we have to make sure a large part of the energy is deposited beyond the critical point in the supersonic region \citep{Leer1980}. Such complex developments will be undertaken in a later study.

We will focus then on the latitudinal distribution of the wind. At minimum of activity, the wind is very organized : at the equator, the magnetic structures being mainly closed loops, the wind tends to be trapped in the corona and is slower ; at higher latitudes, the magnetic field is more open, which allows the wind to reach its maximum speed. As the star approaches maximum of activity, the topology becomes more complex, with more closed loops forming at the surface of the star, which reduces the wind velocity. What is very interesting in our model is that, due to asymmetry, we can clearly see the wind slowing down first in the southern hemisphere only around year 3 after the first minimum, and then in the northern hemisphere around year 4 ; then it speeds up again at high latitudes around year 6 for the southern hemisphere and year 7 for the northern hemisphere. At $20R_\odot$, the time delay of the wind velocity between the northern and southern poles is 5 months. Compared to 9 months for the delay in the magnetic field asymmetry, we can see that the wind tends to smooth the asymmetry. It is difficult to say if this tendency is recovered in observations: in \cite{Tokumaru2015}, the asymmetry in the wind speed is found to be at most a year, with the possibility of being less; for the corresponding cycles, the asymmetry in the sunspot numbers is between one and two years \citep{Svalgaard2013}. So, like in the model, there seems to be a reduction of the north-south asymmetry between the surface and far away in the wind.

When we compare this result with the same figure from \cite{Pinto2011}, we can notice a few differences. Of course, the asymmetry was absent from their model, which means that their latitudinal slow-downs are perfectly synchronized. Also, in their model, the wind seems to return almost immediately to a minimum-like state after the maximum, while in ours, it really takes most of the second half of the cycle to reach an equatorial zone of slow-down which is as narrow as in the minimum state of activity. This feature in our model is actually closer to what is observed in the Sun : solar wind reconstruction using IPS by \cite{Tokumaru2010} and \cite{Sokol2015} display the same behavior, with some kind of relaxing time between the maximum and minimum of activity of 7.5 years, compared to 6 in our model. We can also point out that, as asymmetry is present in the Sun, we can also notice a delay between the latitudinal variations of the wind in \cite{Sokol2015}.

Finally, we can notice that during maximum, the wind velocity is dropping at the poles in the two hemispheres. This has been observed in Ulysses data and has been interpreted as the presence of high-latitude current sheets \citep{Khabarova2017}.

\begin{figure}
\centering
\includegraphics[width=13cm]{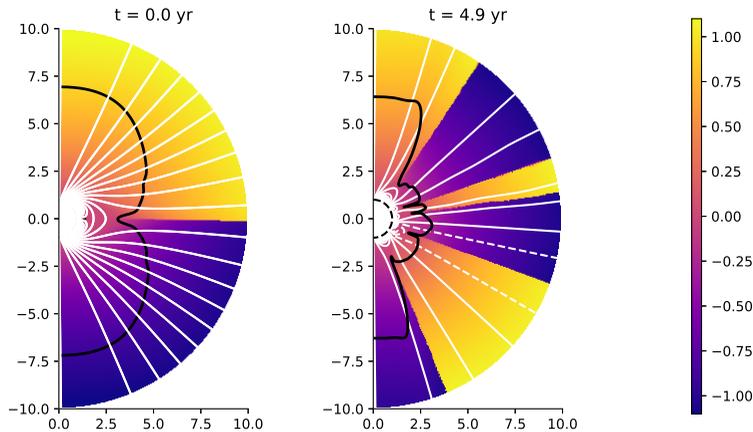}
\caption{Meridional cuts of the Alfvén surface (black line) at minimum and maximum of activity (respectively first and second panel). The color in the background shows the velocity projected on the magnetic field in unit of Mach number. The white lines are the poloidal magnetic field lines of positive polarity in solid and negative polarity in dashed lines. We represent only the 10 first solar radii.}
\label{fig:alfven}
\end{figure}

Figure \ref{fig:alfven} displays the Alfvén surface profile at minimum and maximum (respectively left and right panel). The Alfvén surface corresponds to the surface at which the wind speed becomes faster that the Alfvén speed $v_A = \sqrt{||\boldsymbol{B}||/(4\pi\rho)}$. At minimum, the Alfvén surface is very regular. The Alfvén radius is bigger at the poles than at the equator, due to the magnetic field being stronger. At maximum, we can see the Alfvén surface becoming more irregular due to the change of topology. However at the poles, the Alfvén radius remains pretty large, due to both the magnetic field still being stronger and the drop in the wind velocity observed in Figure \ref{fig:diag_wind} between year 2.5 and 5.5.

We see that our model is qualitatively closer to observations thanks to the asymmetric magnetic field produced by the underlying dynamo model.

\section{Mass and momentum flux}\label{sec:global}

In this section, we will focus on quantities of interest and their time evolution over the cycle in our simulations. The three quantities we are interested in are the average Alfvén radius, the mass loss rate and the angular momentum loss rate. 

\begin{figure}
\centering
\begin{subfigure}[t]{0.65\textwidth}
  \centering
  \includegraphics[width=\textwidth]{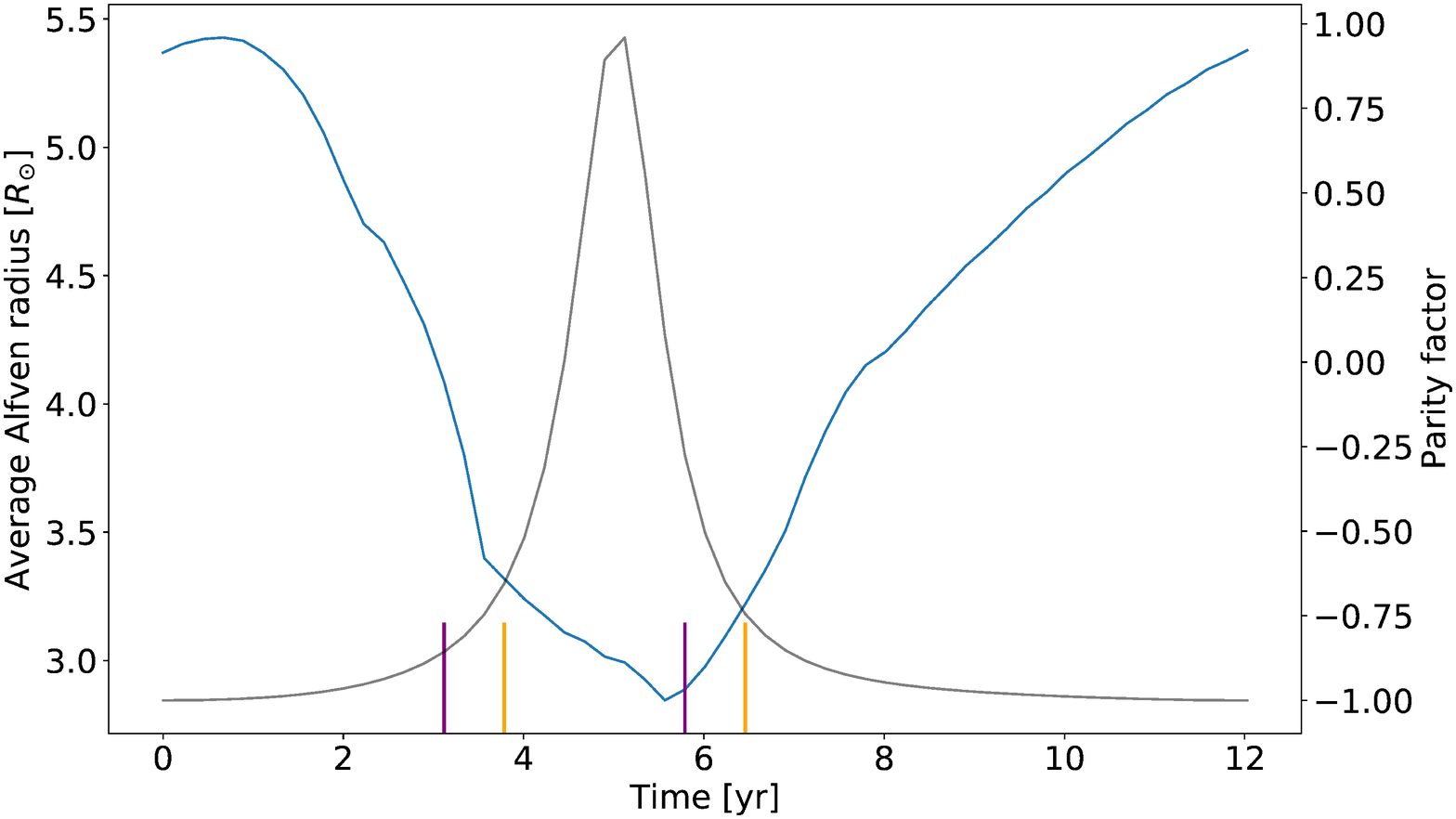}
\end{subfigure}
\begin{subfigure}[t]{0.65\textwidth}
  \centering
  \includegraphics[width=\textwidth]{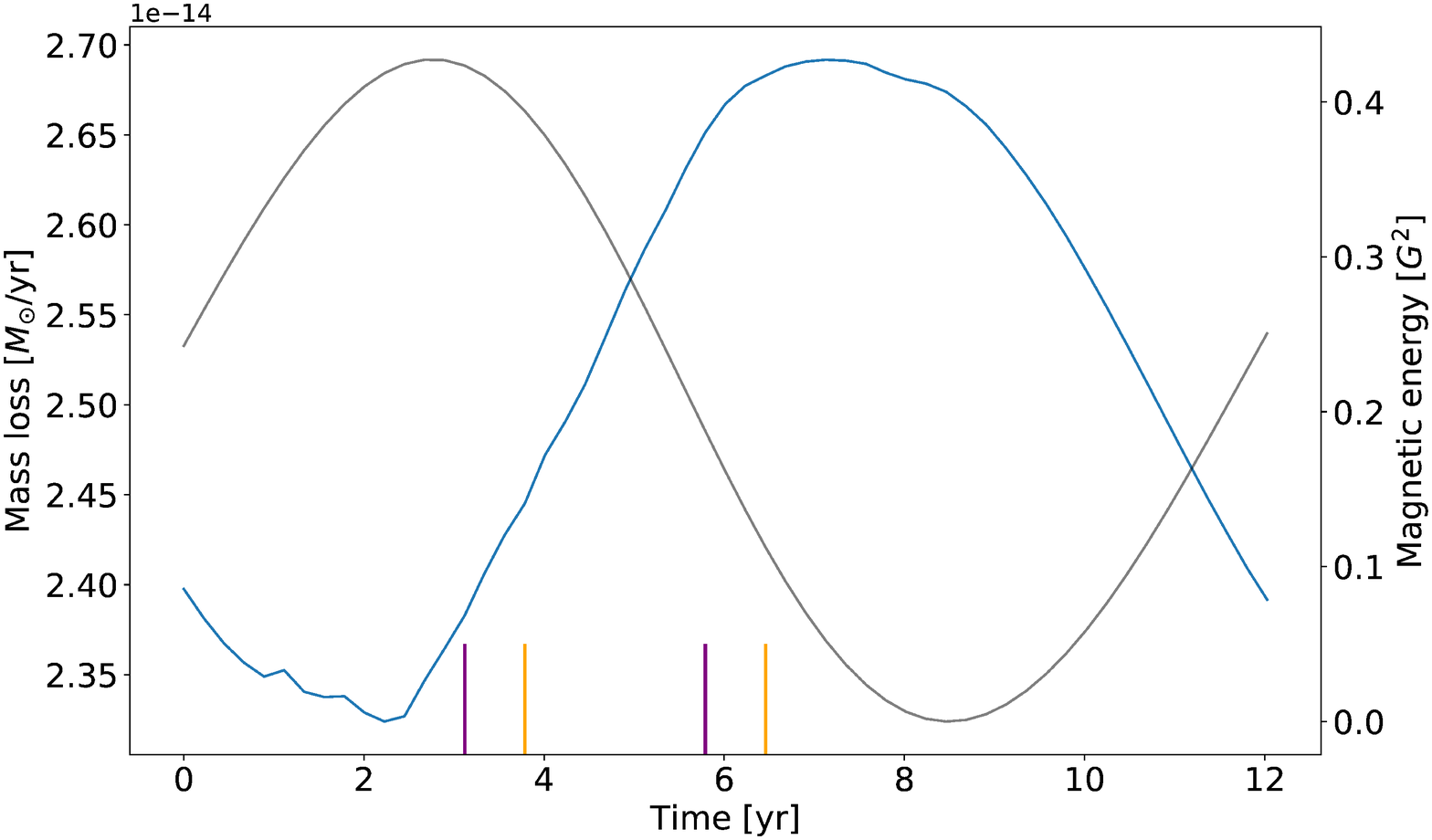}
\end{subfigure}
\begin{subfigure}[t]{0.65\textwidth}
  \centering
  \includegraphics[width=\textwidth]{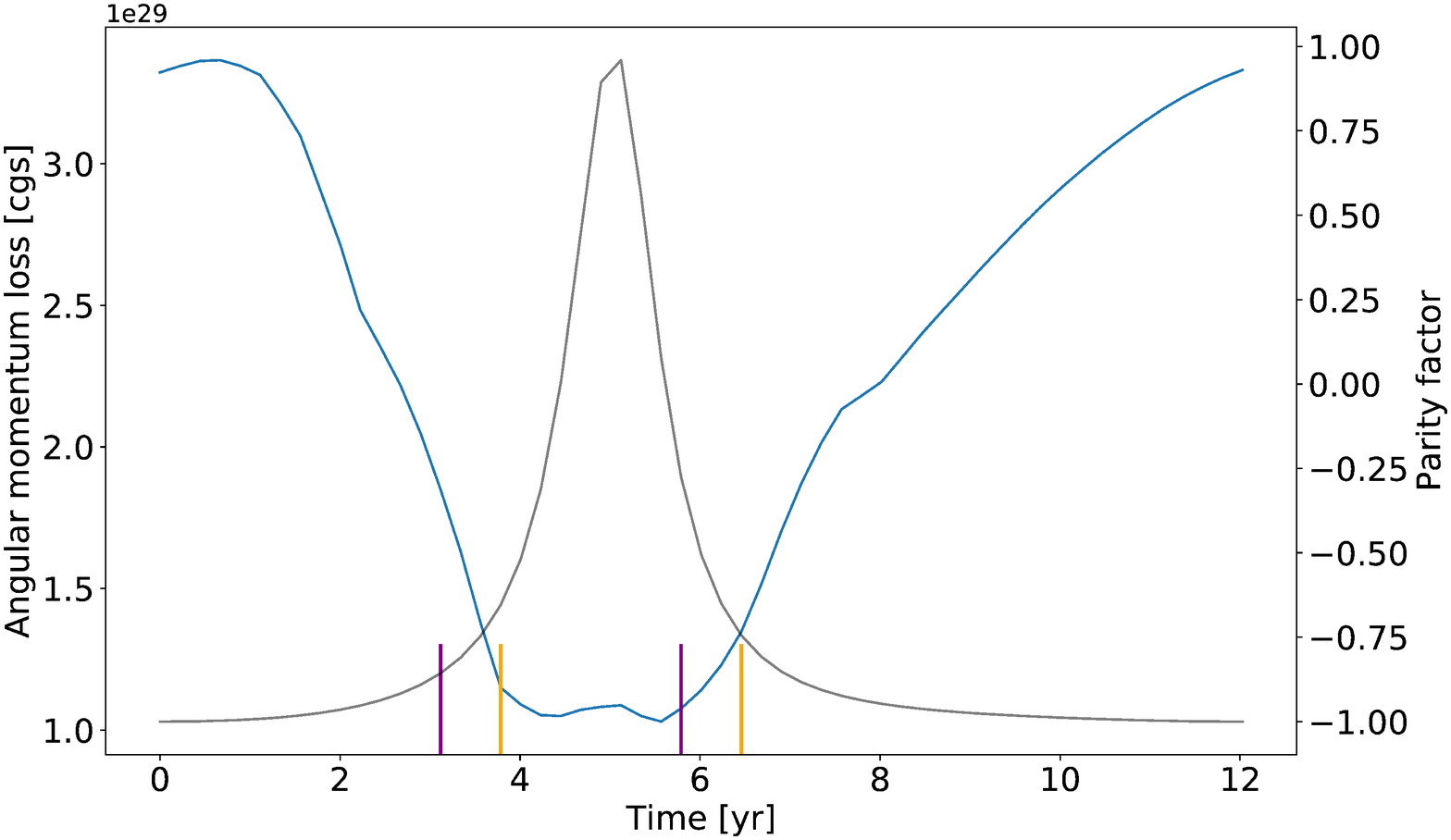}
\end{subfigure}
\caption{Time evolution of several global quantities of interest over an 11-year cycle : average Alfvén radius (first panel), mass loss (second panel) and angular momentum loss (third panel). The purple (orange) ticks indicate the beginning and end of the field reversal in the southern (northern) hemisphere. The grey line is either the parity factor or the surface magnetic energy as indicated in the right y-axis.}
\label{fig:qty_cycle}
\end{figure}

The Alfvén radius $r_A(\theta)$ is defined as the radius at which the wind velocity is equal to the Alfvén velocity. We then define an average Alfvén radius (as in \cite{Pinto2011}) as the average cylindrical radii of the Alfvén surface weighted by the local mass flux $r^2sin\theta\rho\boldsymbol{u}$ crossing the surface of a sphere :
\begin{equation}
\langle r_A \rangle = \frac{\int r^2sin\theta\rho u_r r_A(\theta)d\theta}{\int r^2sin\theta||\rho \boldsymbol{u}||d\theta}.
\label{eq:ra}
\end{equation}

The mass loss rate is defined as :
\begin{equation}
\dot{M} = 2\pi R_0^2\int_0^\pi\rho u_rsin\theta d\theta,
\label{eq:ml}
\end{equation}
where $R_0$ is the radius at which you want to measure the mass loss rate (ie. the radius of the spherical integration surface). In theory, the mass loss rate should be independent of $R_0$. In our case, the mass loss is evaluated at the outer boundary of the domain. 

In the same way, we define the angular momentum loss rate as :
\begin{equation}
\dot{J} = 2\pi R_0^3\int_0^\pi \rho sin^2\theta u_r\left(u_\phi - \frac{B_\phi}{\mu_0\rho} \frac{\boldsymbol{B}_p\cdot\boldsymbol{u}_p}{||\boldsymbol{u}_p||^2}\right)d\theta,
\label{eq:jl}
\end{equation}
where $\boldsymbol{B}_p$ and $\boldsymbol{u}_p$ are respectively the poloidal magnetic field and velocity flow. 

Figure \ref{fig:qty_cycle} shows the evolution of these three quantities over a solar cycle, starting at minimum of activity. Since our wind solution is being relaxed for a given time and a given magnetic field configuration, it means that we can make temporal averages. Note then that the curves we show in Figure \ref{fig:qty_cycle} are averaged in time over 25 reference times, which means that the variations observed are significant compared to temporal variations. We overplotted two relevant quantities to understand our variations : the magnetic surface energy, already defined in equation \eqref{eq:ME}, and the parity factor defined as follows:
\begin{equation}
P = \frac{\alpha_{2,0}^2-\alpha_{1,0}^2}{\alpha_{2,0}^2+\alpha_{1,0}^2},
\end{equation}
where $\alpha_{2,0}^2$ is the energy of the quadrupole and $\alpha_{1,0}^2$ is the energy of the dipole.

The average Alfvén radius goes from 5.5 solar radii at minimum to 3.0 solar radii at maximum. These values are slightly smaller than what was estimated from the angular momentum loss from the Helios mission in \cite{Pizzo1983} and \cite{Marsch1984}, where its largest value was estimated at 12-14 and 13.6-16.6 solar radii respectively. However, when we compare our values to numerical models, the range varies between 2.5 $R_\odot$ and 60 $R_\odot$ : isothermal and polytropic models tend to give a lower estimate for the Alfvén radius \citep{Pneuman1971}. When we compare with \cite{Pinto2011} to see the impact of the Babcock-Leighton model versus the alpha-omega dynamo model, we see that their Alfvén radius goes from 9 $R_\odot$ at minimum to 2.2 $R_\odot$ at maximum. The values are quite similar, it is more the amplitude of the variation over the cycle which has been modified : there is a factor 4 in their case, a factor 2 in our case. When we compare with \cite{Reville2017} for the impact of 2D versus 3D and theoretical dynamo versus observational maps, we see that their Alfvén radius goes from 5 $R_\odot$ at minimum to 7 $R_\odot$ at maximum. The amplitude of the variation is slightly smaller. An explanation proposed for these discrepancies lies in the variations of the dipole component and the correlation with the total magnetic energy, as shown in Figure \ref{fig:mag_energy} : in our case, contrary to \cite{Pinto2011}, we do have an anti-correlation, which is not perfect but does capture most of this effect, which could explain our intermediate results.

\begin{figure}
\centering
\includegraphics[width=13.5cm]{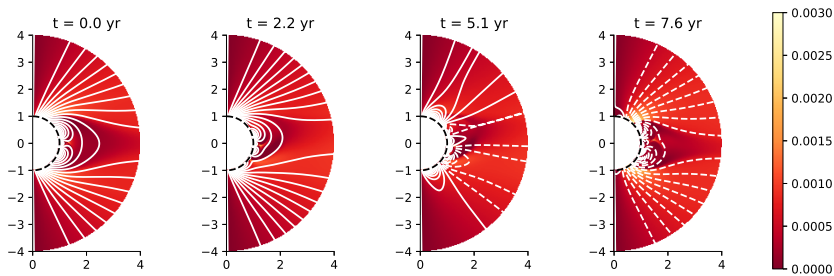}
\caption{Mass flux in the meridional plane at different times in the cycle : the first panel is at $t=0$ year from the first minimum of activity, the second at $t=3.1$ years, the third at $t=4.9$ years and the last at $t=9.5$ years. Bright shades indicate higher mass loss rate. The white lines are magnetic field lines of positive (negative) polarity in solid (dashed) lines. We represent only the 4 first solar radii.}
\label{fig:mas_flux}
\end{figure}

The mass loss we compute varies from 2.35 $10^{-14} \ M_\odot/yr$ to 2.70 $10^{-14} \ M_\odot/yr$ at maximum. From the data in \cite{McComas2008} and the calculations from \cite{Reville2017}, the mass loss inferred by Ulysses reached a minimum value of 2.3 $10^{-14} \ M_\odot/yr$ in 1991 and a maximum value of 3.1 $10^{-14} \ M_\odot/yr$ in 1992-1993. Our minimum mass loss is correct, but the amplitude of the variations is smaller. \cite{Wang1998} indicates that the mass loss rate is minimum at minimum of activity, then rises during the maximum, and then returns to its original value as the star goes back to minimum. Our mass loss has a different behavior : it decreases till it reaches it minimum value around year 3, then rises and reaches its maximum around year 8, before decreasing again until the end of the cycle. This behavior can be explained by the fact that our maximum of polarity and maximum of magnetic energy do not happen at the same time : our mass loss is mostly anti-correlated with the evolution of the magnetic surface energy, which is also the case in the model of \cite{Reville2017}. The variation between the mass loss at minimum and maximum is smaller than in most simulations (13\% in our case, factor 2 in \cite{Pinto2011}, 20\% in \cite{Reville2017}). This small variation of the mass loss may well be a consequence of the delay between the minimum of activity and the minimum defined by topology. We recall also that the geometry of the magnetic field has not been fine-tuned to represent any solar cycle in particular. 

Figure \ref{fig:mas_flux} shows a 2D map of the mass flux in the meridional plane ; this allows us to determine the latitude at which the mass loss is more important. We plotted different relevant moments of the evolution of the mass loss rate : at minimum (year 0), during the local minimal (year 2.2), during the rising phase (year 5.1) and at the maximal value (year 7.6). At minimum of activity, we have a repartition very similar to that of \cite{Pinto2011} : the mass loss comes mainly from the coronal holes located at the poles near the equatorial streamer. However, since the reversals of the two hemisphere happen at different times but still close to one another, we can see that there are less coronal holes than expected that form at year 5.1 (which is almost the multipolar maximum). The minimal value of the mass loss corresponds to a configuration where the polar coronal holes are closing and yet no coronal hole associated to a pseudo-streamer has opened. The maximal value of the mass loss corresponds to a configuration where the polar coronal holes are bigger than at the minimum of activity due to the equatorial streamer which is not completely formed yet.

Finally, our angular momentum loss varies from $3.5 \ 10^{29}$ g cm\textsuperscript{2} s\textsuperscript{-1} at minimum of activity to $1.0 \ 10^{29}$ g cm\textsuperscript{2} s\textsuperscript{-1} at maximum of activity, hence a variation of 70\%. This is the same trend as found in \cite{Pinto2011} or \cite{Reville2015a} : the star loses less angular momentum when it is more multipolar, as shown by the anti-correlation with the parity factor. It can be used to compute the magnetic spin-down timescale, defined as :
\begin{equation}
\delta t_{sd} = \frac{J_\odot}{\dot{J}},
\end{equation}
where $J_\odot$ is the Sun's angular momentum, estimated at $1.84 \ 10^{48}$ g cm\textsuperscript{2} s\textsuperscript{-1} in \cite{Pinto2011} from a 1D seismic solar model \citep{Brun2002}. This yields a magnetic spin-down timescale varying from $1.7 \ 10^{11}$ years to $5.7 \ 10^{11}$ years, which is of the same order of magnitude of what was found in \cite{Pinto2011}.

\begin{figure}
\centering
\begin{subfigure}[t]{0.65\textwidth}
  \centering
  \includegraphics[width=\textwidth]{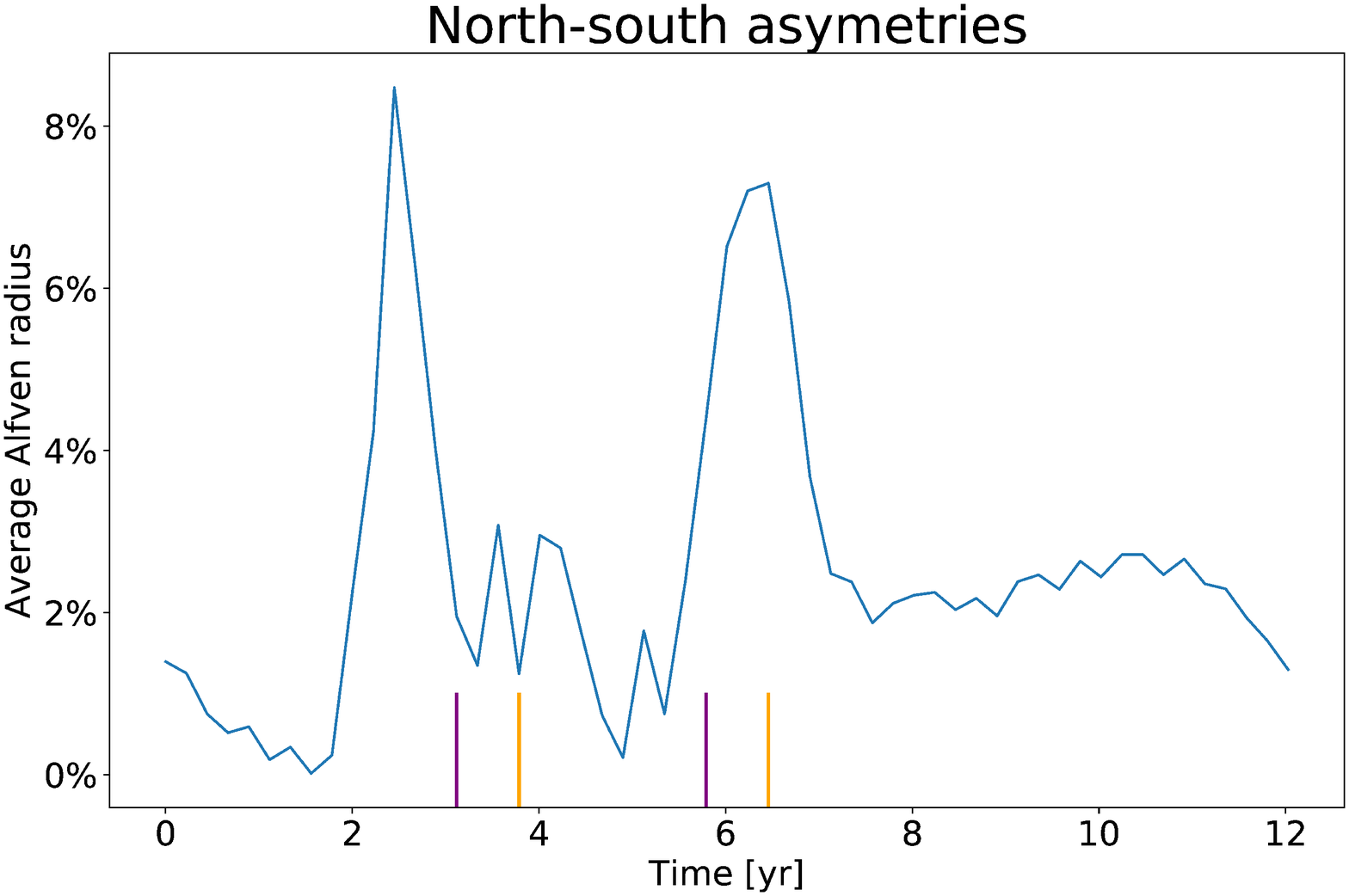}
\end{subfigure}
\begin{subfigure}[t]{0.65\textwidth}
  \centering
  \includegraphics[width=\textwidth]{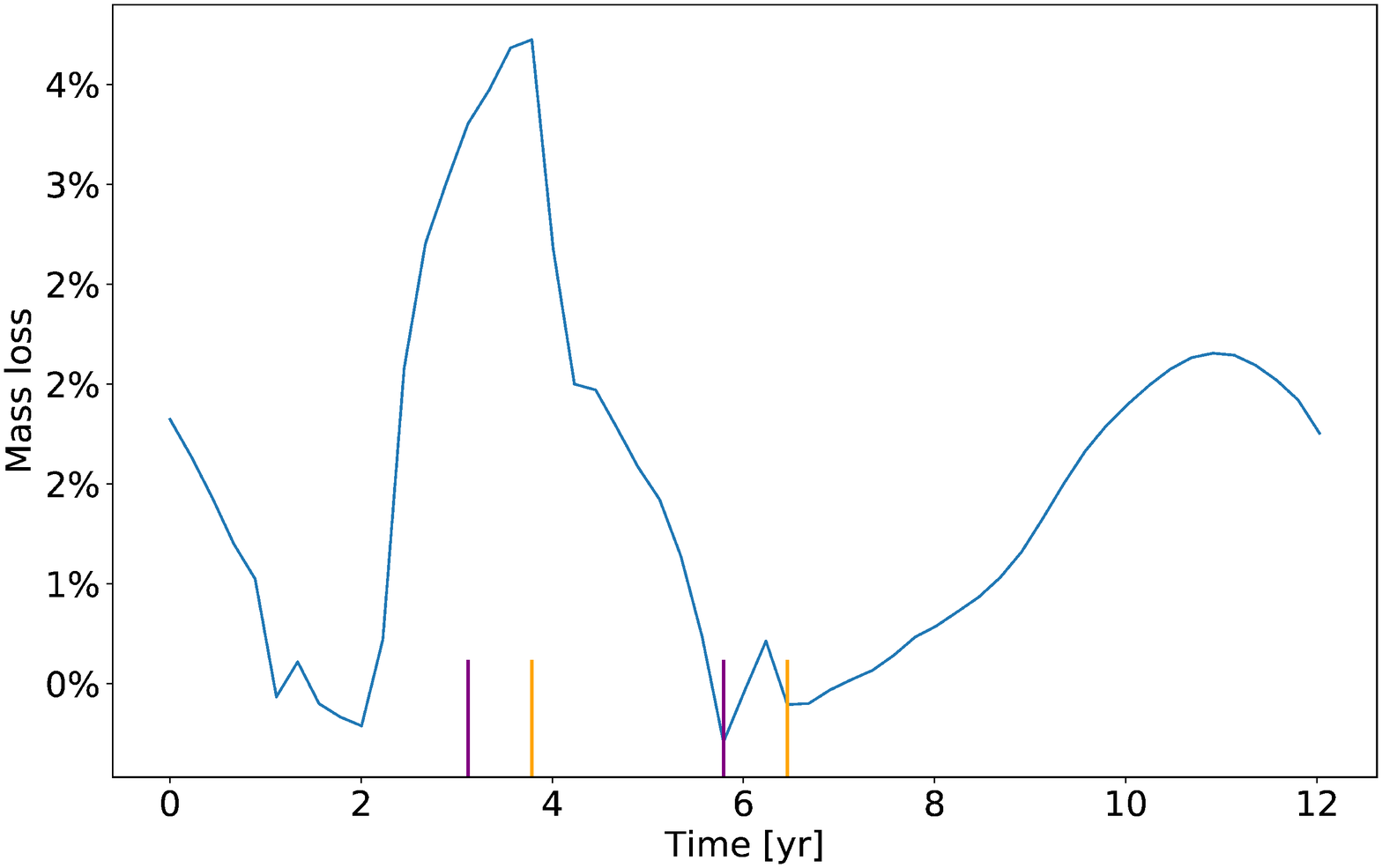}
\end{subfigure}
\begin{subfigure}[t]{0.65\textwidth}
  \centering
  \includegraphics[width=\textwidth]{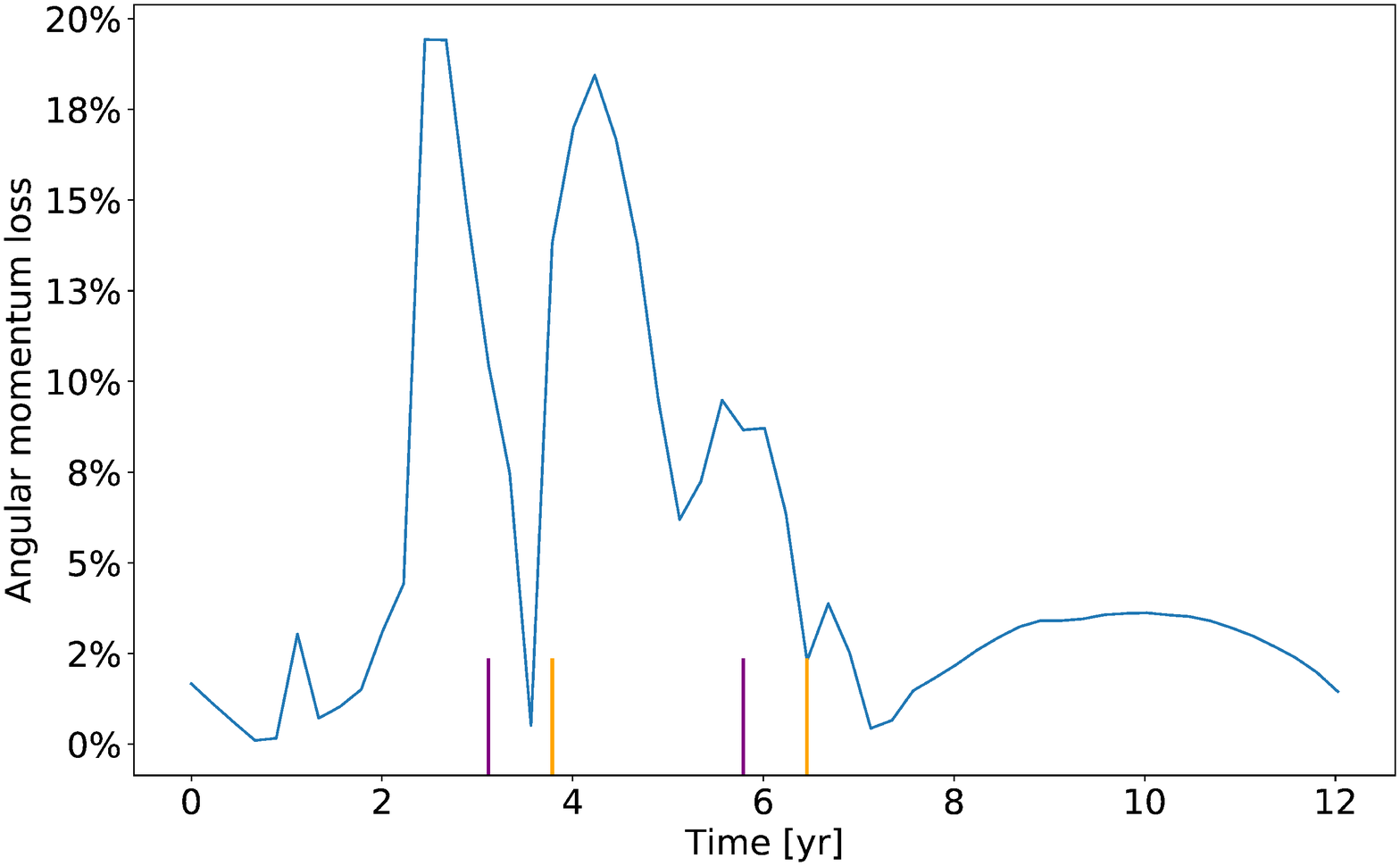}
\end{subfigure}
\caption{Time evolution of the north-south asymmetry over an 11-year cycle for the average Alfvén radius (first panel), mass loss (second panel) and angular momentum loss (third panel). The purple (orange) lines indicate the beginning and end of the field reversal in the southern (northern) hemisphere.}
\label{fig:qty_cycle_asym}
\end{figure}

To finish this section, we study the impact of north-south asymmetry on the variations of these global quantities. In Figure \ref{fig:qty_cycle_asym}, we plot the time evolution of the north-south relative difference for the average Alfvén radius, the mass loss and the angular momentum loss in percentage. To compute these quantities for the northern or southern hemisphere only, we use the same definitions as in \eqref{eq:ra}, \eqref{eq:ml} and \eqref{eq:jl}, except that the integral goes from 0 to $\pi/2$ or from $\pi/2$ to $\pi$. For the average Alfvén radius, the asymmetry goes up to 8\%. It gets bigger around year 2.5 and 6.5, which correspond to right before opening and right after closing of coronal holes of opposite polarities, as shown by the purple (orange) lines for the southern (northern) hemisphere. For the mass loss, due to the fact that $\nabla \cdot \rho \boldsymbol{u} = 0$, we would not expect any asymmetry between the two hemispheres. However, since the wind velocity is influenced by the magnetic field, the asymmetry propagates and is enough to change the mass flux of the hemispheres. The asymmetry is nonetheless less significant that for other quantities, reaching only 3\% at most around year 4, which correspond to opening of coronal holes of opposite polarity. Finally it is for the angular momentum loss that the asymmetry is actually more visible, reaching up to a difference of 20\% between the two hemispheres. Here, the asymmetry is stronger around year 2.5 and year 4.5. To conclude, in these simulations, the asymmetry is also present in global quantities and is more visible at opening and closing of coronal holes of opposite polarity for each hemisphere, which correspond to the reversal of the field. The fact that the asymmetry is stronger during maximum of activity has been observed for the Sun \citep{Tokumaru2015}.

\section{Conclusion and Perspectives}\label{sec:discussion}

We have studied in this work how the properties of the solar dynamo cycle influence its wind and corona. We followed a similar approach as in \cite{Pinto2011}, but went beyond by using more realistic models and considering the effect of north-south asymmetry in the dynamo field. We have used a magnetic field cycle produced by a dynamo code based on a Babcock-Leighton flux-transport model described in \cite{Jouve2007} and including an asymmetry parameter $\epsilon$ as in \cite{DeRosa2012}. This allowed us to recover solar-like features such as the phase-lag between the poloidal and toroidal field or the relative extent of the equatorward and poleward branches seen in the butterfly diagram (Figure \ref{fig:diagpap}). The north-south asymmetry in the magnetic field, that we took into account by introducing an asymmetry in the Babcock-Leighton term, couples the symmetric and antisymmetric family modes of the dynamo, resulting in a delay between the polarity reversals of the northern and southern hemisphere of the star as observed in the Sun. We recall that this model was not parameterized to reproduce a specific solar cycle, our aim was to understand from a theoretical point of view which physical ingredients allow us to recover features similar to observations of the solar corona. We then coupled the output of the dynamo model to a polytropic wind model (note that in \citealt{Pinto2011} the wind was considered isothermal) in spherical geometry \citep[adapted from][]{Reville2015a}. We computed 54 states of relaxed wind over an 11-year dynamo cycle : we studied the influence of a complex magnetic topology with north-south asymmetry on the wind.

We find that an anti-correlation between the energy of the dominant magnetic mode and the total magnetic energy at the surface tends to reduce the variations of the global quantities, as in \citet{Reville2017}. In \cite{Pinto2011}, the Alfvén radius varies by a factor 4  and the mass loss varies by 60\% over the cycle ; in our study, we find that the Alfvén radius varies by a factor 2 and the mass loss by 13\%, which is closer to what was found in \citet{Reville2017} based on magnetograms of the Sun, where the Alfvén radius varies by 30\% and the mass loss by 20\%.

The variations of the different magnetic modes over the cycle induce a complex corona with very different features at minimum and maximum. At minimum we have faster wind at the poles and slower wind at the equator ; the wind is very organized with a main streamer at the equator separating the positive from the negative polarity regions and coronal holes are located only at the poles. At maximum the distribution of the wind speed loses a clear latitudinal dependency, similarly to the Sun as observations with Ulysses clearly demonstrated \citep{McComas2008}. Along the cycle more and more pseudo-streamers and streamers emerge, with coronal holes of opposite polarity opening first at mid-latitudes, first in the southern hemisphere and then in the northern, leading to a complex latitudinal polarity repartition with up to 6 different regions of different polarity next to one another at maximum. The mass loss originates from coronal holes and close to the streamers : as the corona evolves over the dynamo cycle, so does the mass loss in latitude and amplitude. All these elements show that at a given latitude, the dynamics of the corona change drastically over a dynamo cycle.

In the dynamo model studied here, the dipole and quadrupole were not necessarily the strongest mode in their mode family. This effect is particularly visible at the poles, where higher modes tend to dominate the magnetic field. This influences the delay between the minimum of sunspot activity and dipolar activity, resulting in a different mass loss profile as what is usually computed \citep[see, e.g.][]{Pinto2011}. This however allows us to really pinpoint the interplay between the different modes and the solar wind. As shown in Figure \ref{fig:qty_cycle}, the Alfvén radius and the angular momentum loss rate are anti-correlated with the parity factor, because they are highly sensitive to topology, and which also indicates that only the large-scale modes have a significant impact on the lever arm (e.g. Alfvén radius) for the braking of the star. On the other hand in our model, the mass loss rate is much less sensitive to topology, it mainly follows the global magnetic surface energy. We recall that this result may depend on the fact that we used basic heating for the corona, this needs to be confirmed for more realistic solar atmospheres. Nevertheless we stress here the importance of the detailed energy repartition between the scales and the relative phase of the two symmetry families of the dynamo to assess the quality of a given dynamo solution when compared to the Sun.

The asymmetry in the dynamo model also allowed us to recover more realistic profiles for the wind distribution along the cycle : in Figure \ref{fig:diag_wind} we obtain qualitatively the same structure as displayed in \cite{Sokol2015} using IPS data from \cite{Tokumaru2010}, with a time-latitude variation correlated with the dynamo cycle. The asymmetry is clearly visible, with the southern hemisphere wind speed slowing down first in this particular example. The transition between maximum and minimum of activity is as smooth as in the solar observations. It is also worth noting that the north-south asymmetry was a 9-month delay for the dynamo generated field, and translated into a 5-month delay for the wind speed at 0.1 AU. The wind seems thus to smooth the asymmetry.
Some studies indicate even that, given the proper range of parameters, the asymmetry can slightly appear naturally in flux-transport models if there is a non-linear coupling \citep{Shukuya2017}. It could be interesting to do a more systematic study of the influence of the delay between the two hemisphere reversals. In our case, the delay is shorter than in the Sun. The asymmetry is also present in global quantities such as the Alfvén radius, the mass loss and the angular momentum. If this result can be applied to the Sun, it means that we have to be take it into account when inferring theses global quantities from measures taken at a single latitude, especially for the angular momentum loss.

We have to bear in mind the limitations of our model. The fact that the dynamo was not calibrated to be exactly solar-like has for direct consequence that all physical quantities may not necessarily exhibit a solar-like behavior. For instance the polytropic approach is a simple approximation of the heating of the corona, which leads to e.g. a smaller Alfvén radius than the expected solar Alfvén radius. We could have calibrated our wind model in order to increase it, but this would have also modified other physical quantities. Instead, we chose to calibrate our model to reproduce the mean mass loss rate of the Sun for it is a better known value deduced from in situ observations. Finally we recall that with such a quasi-static approach, we can understand some of the influence of the dynamo on the wind, but we cannot include the influence of the wind on the dynamo, if any. Still our results show interesting relation between dynamo and wind properties.

For the future, several aspects remain to be explored. A more realistic coronal heating such as the ones prescribed in \cite{Cranmer2007} or \cite{Riley2015} should be use to improve the realism of the wind solutions. \cite{Reville2017} have shown that considering the full 3D topology of the solar magnetic field also brings a lot more information and can help recovering more realistic features of the solar environment. For example, an important feature when comparing to observational data is the tilt of the heliospheric current sheet, which is omitted here due to the assumption of axisymmetry. We are currently developing a 3D set-up up to 1 AU to look at such tendencies. The north-south asymmetry we studied was also not variable in time, which should be the case in the Sun: we could add stochastic noise in the future to make it even more realistic on longer timescales and thus obtain different delays depending on the cycle. Please note that the two codes used can also be used to model other stars than the Sun, a similar study could be performed for any star on the main sequence. Finally, the best way to understand the full interplay between the dynamo and the wind would be to consider a dynamical coupling, to avoid any non-causal perturbations, which we intend to explore in a future work, along with some of the improvements listed above.

\section*{Acknowledgments}

We thank R. Pinto for useful discussions. This work was supported by a CEA "Thèse Phare" grant, by CNRS and INSU/PNST program and by CNES SHM funds. Computations were carried out using CEA CCRT and CNRS IDRIS facilities within the GENCI 20410133 allocation. 

\bibliographystyle{jpp}
\bibliography{article}

\end{document}